\documentclass[aps,prb,reprint,superscriptaddress]{revtex4-2}
\usepackage[utf8]{inputenc}

\usepackage[hidelinks]{hyperref}

\usepackage{hyperref}
\usepackage{color}
\hypersetup{colorlinks=true}
\usepackage{graphicx}
\usepackage{bm}
\usepackage{amsmath}
\usepackage{amssymb}
\usepackage{xspace}

\usepackage{physics}
\usepackage{siunitx}
\usepackage{chemformula}

\usepackage[]{todonotes}
\newcommand{\dip}{\Omega_{\mathrm{d}}}
\newcommand{\PT}{$\mathcal{PT}$}
\newcommand{\hc}{\mathrm{h.c.}}
\newcommand{\NN}{N_{R}}
\newcommand{\tend}{t_{\mathrm{end}}}
\newcommand{\fj}{f_J}
\newcommand{\fg}{f_G}
\newcommand{\fsto}{f_{\mathrm{STO}}}
\newcommand{\fdip}{f_{\mathrm{dip}}}

\graphicspath{{./figures/}}
\begin{document}
	
\title{Non-linear dynamics of the non-Hermitian Su-Schrieffer-Heeger model}

\author{Pieter M. Gunnink}
\email{p.m.gunnink@uu.nl}
\affiliation{Institute for Theoretical Physics and Center for Extreme Matter and Emergent Phenomena, Utrecht University, Leuvenlaan 4, 3584 CE Utrecht, The Netherlands}

\author{Benedetta Flebus}
\affiliation{Department of Physics, Boston College, 140 Commonwealth Avenue, Chestnut Hill, Massachusetts 02467, USA}

\author{Hilary M. Hurst}
\affiliation{Department of Physics and Astronomy, San Jos\'{e} State University, San Jos\'{e}, California, 95192, USA}

\author{Rembert A. Duine}
\affiliation{Institute for Theoretical Physics and Center for Extreme Matter and Emergent Phenomena, Utrecht University, Leuvenlaan 4, 3584 CE Utrecht, The Netherlands}
\affiliation{Department of Applied Physics, Eindhoven University of Technology, P.O. Box 513, 5600 MB Eindhoven, The Netherlands}

\date{\today}
\begin{abstract}
	We numerically determine the robustness of the lasing edge modes in a spin-torque oscillator array that realizes the non-Hermitian Su-Schrieffer-Heeger model. Previous studies found that the linearized dynamics can enter a topological regime in which the edge mode is driven into auto-oscillation, while the bulk dynamics are suppressed. Here we investigate the full non-linear and finite-temperature dynamics, whose understanding is essential for spin-torque oscillators-based applications. Our analysis shows that the lasing edge mode dynamics persist in the non-linear domain for a broad range of parameters and temperatures. 	
	We investigate the effects of perturbations relevant to experimental implementations and discuss which ones might be  detrimental to the stability of the lasing edge mode. 	
	Finally, we map our model onto a photonic model. Our analysis has the potential to shed light onto the dynamics of a plethora of non-Hermitian systems with non-linearities. 	
\end{abstract}
\maketitle

\section{Introduction}
The application of topology to condensed matter systems has been profoundly fruitful on both theoretical and experimental fronts and has lead to the discovery of a wide range of new phenomena and materials \cite{hasanColloquiumTopologicalInsulators2010a}. Recently, considerable effort has been devoted towards the exploration of non-Hermitian systems \cite{benderComplexExtensionQuantum2002,benderRealSpectraNonHermitian1998} with active gain and loss. A framework for addressing non-Hermitian topological phases has  been provided by the growing field of topological theories of non-Hermitian systems \cite{bergholtzExceptionalTopologyNonHermitian2021, kawabataSymmetryTopologyNonHermitian2019,gongTopologicalPhasesNonHermitian2018a,shenTopologicalBandTheory2018}. 

Recent works have shown that the bulk-boundary correspondence \cite{chiuClassificationTopologicalQuantum2016}, which is the cornerstone of topology in Hermitian systems, also holds for specific non-Hermitian systems \cite{kawabataAnomalousHelicalEdge2018}, although not in general \cite{martinezalvarezNonHermitianRobustEdge2018,borgniaNonHermitianBoundaryModes2020a, yaoEdgeStatesTopological2018}. Some systems also exhibit a non-Hermitian skin effect, where even the bulk modes can be very sensitive to boundary conditions \cite{kunstBiorthogonalBulkBoundaryCorrespondence2018}. Nevertheless, the existence of non-Hermitian edge modes has been shown in a variety of systems, such as microring resonators \cite{partoEdgeModeLasing1D2018, bandresTopologicalInsulatorLaser2018,harariTopologicalInsulatorLaser2018a} and electrical circuits \cite{albertTopologicalPropertiesLinear2015,ezawaNonHermitianHigherorderTopological2019,liuGainLossInducedTopological2020}.

One of the most striking properties of non-Hermitian topological phases is the co-existence of lasing edge modes, i.e. edge states with gain-like dynamics, with a purely real bulk spectrum. Most importantly, these modes are topologically protected \cite{kawabataSymmetryTopologyNonHermitian2019} and can therefore be useful in applications, since they are robust against disorder. In photonics these lasing modes have been used to build a single-mode laser stable against perturbations \cite{zhaoTopologicalHybridSilicon2018}. 

The majority of developments in non-Hermitian topological insulators have been in the field of photonics \cite{ozawaTopologicalPhotonics2019,martinezalvarezTopologicalStatesNonHermitian2018, ozdemirParityTimeSymmetry2019,suDirectMeasurementNonHermitian2021}, where gain and dissipation can be readily tuned.
Recent works have also unveiled non-Hermitian topological phases in mechanical \cite{ghatakObservationNonHermitianTopology2020, scheibnerNonHermitianBandTopology2020}, electrical \cite{albertTopologicalPropertiesLinear2015,ezawaNonHermitianHigherorderTopological2019,liuGainLossInducedTopological2020, helbigGeneralizedBulkBoundary2020} and magnetic systems \cite{leeMacroscopicMagneticStructures2015,mcclartyNonHermitianTopologySpontaneous2019,dengNonHermitianSkinEffect2021}. Here we focus on magnetic systems, in which the loss is inherently present due to coupling of the magnonic excitations to the lattice and whose dynamics can be driven using spin-transfer torques. Due to the tunability of gain and loss, magnetic systems might represent a nearly ideal system to explore non-Hermitian phenomena.

Specifically, we consider the topology of the one-dimensional (1D) array of spin-torque oscillators (STOs) as shown in Fig.~\ref{fig:setup}, building on the work of Flebus \textit{et al.} \cite{flebusNonHermitianTopologyOnedimensional2020a}. Spin-torque oscillators are current-driven magnetic nanopillars, whose magnetization dynamics are determined by the balance of spin current injection and intrinsic (Gilbert-like) dissipation \cite{slavinNonlinearAutoOscillatorTheory2009}. It has been experimentally shown that the coupling between STOs arranged in an array can be tuned \cite{kakaMutualPhaselockingMicrowave2005,mancoffPhaselockingDoublepointcontactSpintransfer2005}. Flebus \textit{et al.} \cite{flebusNonHermitianTopologyOnedimensional2020a} have shown that, by modulating the coupling between STOs and the local spin injection, the array can be driven into the topological phase of the non-Hermitian Su-Schrieffer-Heeger (SSH) model \cite{suSolitonsPolyacetylene1979}, known to host lasing edge states \cite{lieuTopologicalPhasesNonHermitian2018, yokomizoNonBlochBandTheory2019}. However, STOs also exhibit strong non-linear effects, such as a non-linear frequency shift \cite{slavinNonlinearAutoOscillatorTheory2009}. 
Furthermore, thermal fluctuations have also been shown to introduce significant noise into these systems \cite{tiberkevichMicrowavePowerGenerated2007}. 
In this work, we aim to investigate in detail how non-linearities and thermal fluctuations affect the topological character of a 1D array of STOs, in order to assess the experimental feasibility of this setup. Our results can be straightforwardly generalized to the realization of this model in photonic systems, which also exhibit non-linear and stochastic dynamics, making our work of interest to a broader audience.

This work is organized as follows: in Sec.~\ref{sec:model} we start by introducing our model and we analyze its topological properties. In Sec.~\ref{sec:simulations}, we discuss in detail the numerical simulations deployed to investigate the non-linear dynamics at finite temperatures. 
We present the results of our simulations for a wide range of parameters in Sec.~\ref{sec:results} and we identify the parameter regions where the lasing edge mode is realized. In Sec.~\ref{sec:photonic} we show how our model is similar to previous implementations of the non-Hermitian SSH model in photonic systems. A summary and conclusion are given in Sec.~\ref{sec:conclusion}. Finally, we discuss the details of our numerical and analytical calculations in, respectively, Appendix~\ref{app:method} and \ref{app:ham}.

\section{System}
\label{sec:model}
\begin{figure}
	\includegraphics[width=\columnwidth]{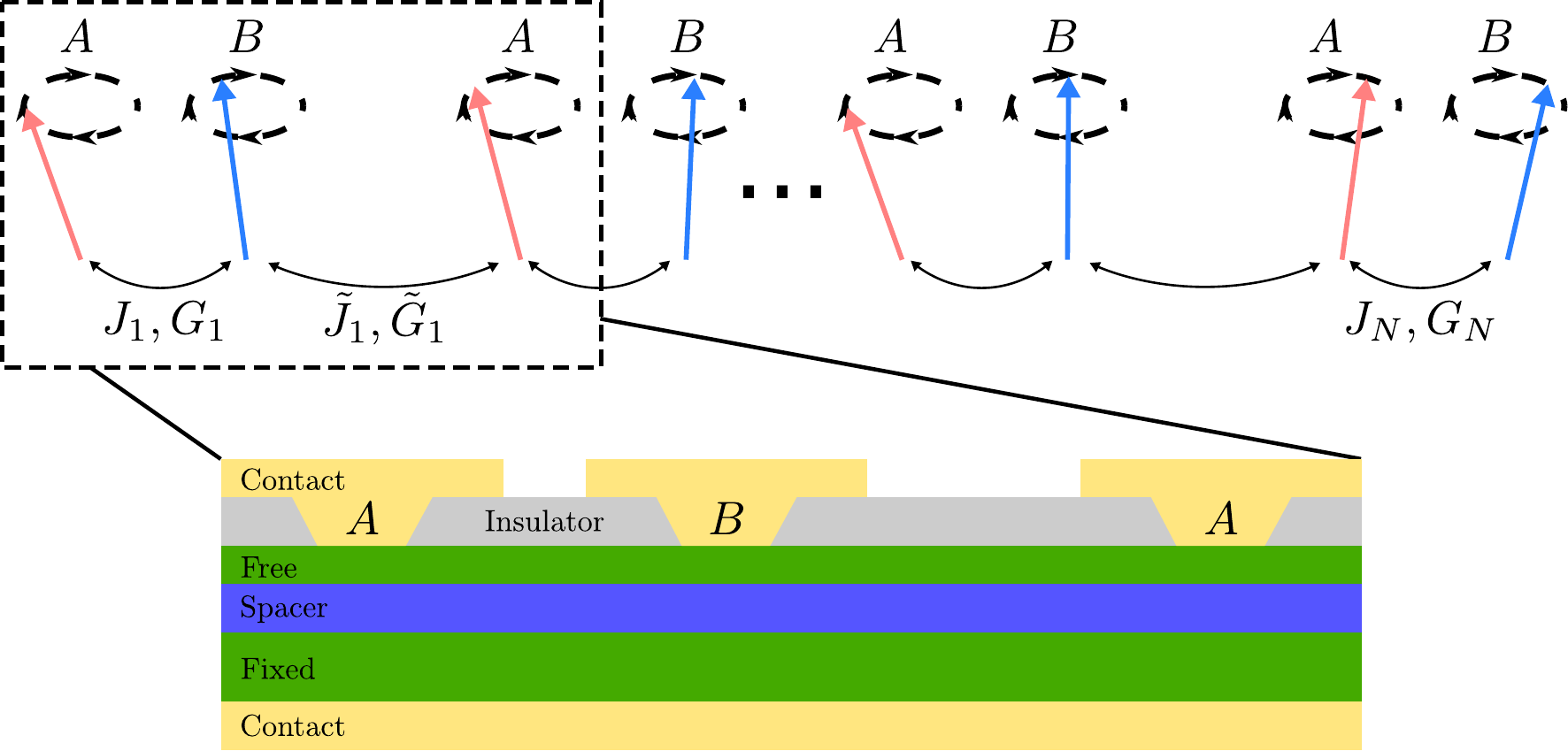}
	\caption{
		Schematic illustration of the model considered in this work. The array of STOs is represented as a 1D lattice with a two-sublattice (A and B) unit cell. The STOs are connected via a metallic spacer that mediates an intra-cell ($J_i,G_i$) and inter-cell ($\tilde{J}_i, \tilde{G}_i$) complex nearest-neighbor hopping. The inset shows a possible experimental realization of the STOs array, with the stacking of contact\textbar fixed layer\textbar spacer\textbar free layer\textbar contact. The magnetic layers extend throughout the array, and spin waves excited by one STO can reach and dissipatively couple the neighboring STOs.
		\label{fig:setup}}
\end{figure}

We consider an array of $2N$ STOs that realize the non-Hermitian SSH model, as shown in Fig.~\ref{fig:setup}. A STO consists of a magnetic polarizing layer separated from a magnetic free layer by a thin spacer. An external magnetic field $\bm{H}=H_0 \hat{\bm{z}}$ sets the equilibrium direction of the magnetic order parameter $\bm{m}$ of the free layer. The polarizing layer converts a DC current into a spin current $J_s$, which, in turn, exerts a spin-transfer torque on the magnetic order parameter $\bm{m}$. The loss and gain dynamics of the ferromagnetic order parameter $\bm{m}$ associated with each nanopillar is described by the Landau-Lifshitz-Gilbert (LLG) equation as \cite{slavinNonlinearAutoOscillatorTheory2009}

\begin{equation}
	\begin{split}
	\partial_t \bm{m}_{\eta,i}|_0=\omega_{\eta,i} \bm{\hat{z}} \times \bm{m}_{\eta,i} +\alpha_{\eta,i} \bm{m}_{\eta,i} \times \partial_t \bm{m}_{\eta,i} \\
	+ J_{s\eta,i} \bm{m}_{\eta,i} \times (\bm{m}_{\eta,i} \times \hat{\bm{z}}),
	\end{split}
	\label{eq:single-sto}
\end{equation}
where $i$ labels the unit cells and $\eta=A,B$. Here, $\omega_{\eta,i}=\gamma_{\eta,i}(H_0-4\pi m_{z;	\eta,i})$ is the ferromagnetic resonance frequency, $\gamma_{\eta,i}$ is the gyromagnetic ratio and $M_{\eta,i}$ the saturation magnetization. $\alpha_{\eta,i} \ll 1$ is the Gilbert damping parameter that captures the relaxation of the macrospin Kittel mode. The last term is the spin-transfer torque exerted by the spin current $J_{s\eta,i}$ on the magnetic order parameter. 

We consider three kinds of intra (inter)-cell couplings between the STOs. Firstly, we account for the Ruderman-Kittel-Kasuya-Yosida (RKKY)-type exchange, parameterized by the frequencies $J_i $ $(\tilde{J}_i)$. Secondly, there is a dissipative coupling $G_i  $ $(\tilde{G}_i)$, which is mediated by spin pumping through the spacer layers \cite{houshangSpinwavebeamDrivenSynchronization2016}. The exchange and dissipative couplings only couple the nearest neighbors, as indicated in Fig.~\ref{fig:setup}. These two coupling were already introduced in Ref.~\cite{flebusNonHermitianTopologyOnedimensional2020a}. Additionally, in this work we also introduce the dipolar couplings between the STOs, which affect the non-linear dynamics \cite{locatelliEfficientSynchronizationDipolarly2015,slavinTheoryMutualPhase2006}.

The dynamics of the coupled array are then described by

\begin{gather}
	\begin{split}
\partial_t \bm{m}_{A,i}|_{\mathrm{coup}} = -\bm{m}_{A,i} \times \left( J_i \bm{m}_{B,i} + \tilde{J}_{i-1} \bm{m}_{B,i-1}\right)  \\
\hspace{1em} - G_i \bm{m}_{B,i}  \times \partial_t \bm{m}_{B,i}  
- \tilde{G}_{i-1}  \bm{m}_{B,i-1}  \times \partial_t \bm{m}_{B,i-1}  \\
\shoveleft{-\dip\bm{m}_{A,i} \times \sideset{}{'}\sum_{\eta,j}  \frac{3\hat{\bm{x}} \left(\hat{\bm{x}}\cdot\bm{m}_{\eta,j}\right) -\bm{m}_{\eta,j} }{r_{A\eta,i j}^3},}  \label{eq:sto-coupled-A}\\
	\end{split}\\
	\begin{split}
		\partial_t \bm{m}_{B,i}|_{\mathrm{coup}} = -\bm{m}_{B,i} \times \left( J_i \bm{m}_{A,i} + \tilde{J}_{i} \bm{m}_{A,i+1}\right)  \\
		\hspace{1em} - G_i \bm{m}_{A,i}  \times \partial_t \bm{m}_{A,i}
		- \tilde{G}_{i} \bm{m}_{A,i+1}  \times \partial_t \bm{m}_{A,i+1}  \\
		\shoveleft{-\dip\bm{m}_{B,i} \times \sideset{}{'}\sum_{\eta,j}  \frac{3\hat{\bm{x}} \left(\hat{\bm{x}}\cdot\bm{m}_{\eta,j}\right) -\bm{m}_{\eta,j} }{r_{B\eta,i j}^3},} \label{eq:sto-coupled-B}
	\end{split}
\end{gather}
where $\sum\nolimits'$ indicates that the sum excludes the self-interaction. The dipolar interaction is parametrized by $\dip=\gamma_{\eta,i}  V_{\mathrm{eff}\eta,i}/a^3$, where $V_{\mathrm{eff}\eta,i}$ is the effective volume of a STO and $a$ is the separation distance between STOs, which we assume to be constant. $r_{\eta \eta,i' j}=R_{\eta \eta,i' j}/a$ is the normalized distance between $\mathrm{STO}_{\eta i}$ and $\mathrm{STO}_{\eta' j}$. Here we have incorporated terms proportional to the Gilbert-like on-site damping of the $G_i,\tilde{G}_i$ couplings into the Gilbert damping, i.e. $\alpha_{A,i}=\alpha_{\eta,i}+G_i+\tilde{G}_{i}$ and $\alpha_{B,i}=\alpha_{\eta,i}+G_i+\tilde{G}_{i-1}$.
In what follows we assume identical unit cells and drop the dependency on the unit cell index $i$, unless stated otherwise, such that $\omega_{\eta,i}=\omega$, $\alpha_{\eta,i}=\alpha+G+\tilde{G}$, $J_i=J, \tilde{J}_i=\tilde{J}$, $G_i=G$ and $\tilde{G}_i=\tilde{G}$. Furthermore, we assume spin-current injection only on sublattice A, i.e. $J_{s,B}=0, J_{s,A}=J_s$. 

The equations of motion for this system are then given by Eqs.~(\ref{eq:single-sto}-\ref{eq:sto-coupled-B}), which can be linearized around the equilibrium direction of the magnetic order parameter.
We write $\bm{m}_{\eta,i}=(m^x_{\eta,i},m^y_{\eta,i},1)$ and neglect terms that are second order in the fluctuations from equilibrium. We then introduce the complex variable $2m_{\eta,i}=m^x_{\eta,i}-im^y_{\eta,i}$ and invoke the Holstein-Primakoff transformation \cite{holsteinFieldDependenceIntrinsic1940} $m_{\eta,i}(t)=\langle \eta_i \rangle e^{ -i\omega t}$, where $\eta_{i}=a_i,b_i$ are second-quantized operators annihilating magnons at sublattice $\eta$ and obeying bosonic commutation relations. From the corresponding Heisenberg equation of motion we can identify the effective quadratic magnon Hamiltonian, which we will use next for topological classification. 

The non-linear character of the equations of motion means that the Hamiltonian should also contain higher order interactions. The on-site dynamics, i.e. Eq.~(\ref{eq:single-sto}), would introduce on-site interaction terms, and the exchange interaction would result in quartic and higher order interactions terms between neighboring spins. How these interaction terms affect the topology is still an unanswered question \cite{ezawaNonlinearityinducedTransitionNonlinear2021, ezawaNonlinearNonHermitianHigherorder2022}. We therefore only consider the quadratic Hamiltonian to determine the topological properties. However, in our numerical simulations we do take into account the non-linearities of the equations of motion, thus capturing the full	 dynamics resulting from the higher order interaction terms.

Our starting point is the $\mathcal{PT}$ (parity-time) symmetric case, analyzed as well by Flebus \textit{et al.} \cite{flebusNonHermitianTopologyOnedimensional2020a}. Here $J_s=2\alpha\omega$ and $\dip=G=\tilde{G}=0$. In this regime, the system hosts two edge modes with energies $\Re E - \omega=0$ and $\Im E \neq 0$ for $|J|<|\tilde{J}|$. The \PT-symmetry indicates the system is invariant under combined parity (swapping site A with B and vice versa) and time reversal ($t\rightarrow -t$) operations \cite{benderComplexExtensionQuantum2002,benderRealSpectraNonHermitian1998}.  We assume the strength of the dissipative coupling and the dipole-dipole interactions to be small compared to $\omega$, and treat them as perturbations. We note here that we do have access to the full (including non-linear) dynamics that result from the dissipative and dipolar coupling and only treat them as perturbations in the topological analysis. In the simulations that follow we describe the full dynamics of the system, including dissipative and dipolar couplings. 

The Hamiltonian of the \PT-symmetric model is
\begin{equation}
	\begin{split}
	H_i &= \omega \left[a_i^\dagger a_i + b_i^\dagger b_i\right] + i \left( J_{sA}-\alpha\omega \right) a_i^\dagger a_i \\
	&\quad - i\alpha\omega b_i^\dagger b_i -J\left[ a_i^\dagger b_i + \hc \right] \\
	&\quad - \tilde{J}\left[ a_i^\dagger b_{i-1} + \hc \right],
	\end{split}
	\label{eq:ham}
\end{equation}
for $i\neq 1,N$, with open boundary conditions
\begin{equation}
	\begin{split}
	H_j &= \omega \left[a_j^\dagger a_j + b_j^\dagger b_j\right] + i \left( J_{sA}-\alpha\omega \right) a_j^\dagger a_j \\
	&\quad - i\alpha\omega b_j^\dagger b_j  - J\left[ a_j^\dagger b_j + \hc \right] \\
	&\quad - \tilde{J} \left[b_j^\dagger a_l + \hc \right],
		\end{split}
	\label{eq:ham-bc}
\end{equation}
where $j=1,N$ and $l=2,N-1$. The full Hamiltonian, including the dissipative and dipole-dipole coupling, is given in Appendix~\ref{app:ham}.

We first briefly discuss the phase diagram, which captures the linear dynamics. For a full discussion the reader is referred to the earlier work of Flebus \textit{et al.} \cite{flebusNonHermitianTopologyOnedimensional2020a}.
The topological nature of the edge modes can be characterized by a global complex Berry phase  \cite{lieuTopologicalPhasesNonHermitian2018}, i.e., an integer that predicts the number of pairs of  edge modes. The complex Berry phase can be found to be one for $|J|<|\tilde{J}|$, signaling the presence of topologically protected edge states. 

Furthermore, the system has an exceptional point at $|\tilde{J}\pm\alpha\omega|=|J|$, where the system transitions from the \PT-unbroken into the \PT-broken regime \cite{lieuTopologicalPhasesNonHermitian2018}.  In the \PT-unbroken regime the edge state spectra come as complex-conjugated pairs, while the the bulk spectrum is purely real. Thus, the edge mode with positive imaginary energy starts lasing, while the bulk modes remain inactive. In the \PT-broken regime the bulk modes also become complex valued, such that they also will start lasing spontaneously. In order to isolate the dynamics of the lasing edge mode we therefore require $|J|<|\tilde{J}-\alpha\omega|$, i.e. to be in the \PT-unbroken regime.

When dissipative couplings $G,\tilde{G}$ are present, all bulk modes will have a non-zero imaginary component, since the system is no longer \PT-symmetric. The edge mode is still well defined and separated in energy from the bulk modes. However, because all bulk modes have a non-zero imaginary component, these modes can start lasing as well, as was also noted by Flebus \textit{et al.} \cite{flebusNonHermitianTopologyOnedimensional2020a}. We note that chiral-inversion (CI) symmetry protects the stability of the edge states \cite{jinBulkboundaryCorrespondenceNonHermitian2019}, such that the topologically protected edge modes are now present for $|J-iG\omega|<|\tilde{J}-i\tilde{G}\omega|$. Since in almost all experimental realizations of the setup discussed here the dissipative coupling will be much weaker than the RKKY-type coupling, the system will most likely still be in the topologically non-trivial regime. 

For the dipole-dipole interactions we note that the dipolar fields are \PT-invariant, and thus the bulk spectrum will remain real. However, long-range interactions are typically not captured by topological classifications \cite{gongTopologicalPhasesLongrange2016}, and it is unclear from the linearized model alone how the long-range dipolar interaction will affect the lasing edge modes. This will be investigated numerically in the next section.

As was noted before, the phase diagram only captures the linear behavior of the STO array. In the STO array considered here, the STOs are easily driven into the non-linear regime \cite{flebusNonHermitianTopologyOnedimensional2020a,kimLineShapeDistortion2008a}. Non-linearities therefore need to be taken into account when describing this topological array. Thus, we proceed to investigate the full non-linear dynamics numerically.

\section{Simulations}
\label{sec:simulations}
We numerically simulate the system described by Eqs.~(\ref{eq:single-sto}-\ref{eq:sto-coupled-B}), using the parametrization outlined in Appendix~\ref{app:method}. This parametrization maps the magnetic order parameter $m_{\eta,i}$ to the microwave power $p_{\eta,i}$ (which is experimentally measurable) and the azimuthal angle $\phi_{\eta,i}$. The thermal fluctuations are taken into account by using a stochastic field, the strength of which is chosen such that an isolated STO reaches thermal equilibrium \cite{tiberkevichMicrowavePowerGenerated2007}. We note that this noise will  equilibrate the whole array to $2N$ individual STOs in thermal equilibrium, since the couplings between STOs are not taken into account in the equilibration. However, we assume couplings to be weak compared to the on-site dynamics (i.e. $J,\tilde{J}\ll \omega$), making this a valid approximation. The noise is thus chosen to have zero mean and a second-order correlator
\begin{equation}
\begin{gathered}
	\langle f_{\eta,i}(t)f_{\eta',j}(t') \rangle = 0;\\ 
	\langle f_{\eta,i}(t) f_{\eta',j}^*(t') \rangle = 2\delta_{i,j}\delta_{\eta,\eta'} D_{\eta,i}\left(p_{\eta,i}\right) \delta\left(t-t'\right),
\end{gathered}
\end{equation}
where $D_{\eta,i}(p)$ is a diffusion coefficient that characterizes the noise amplitude, which has to be taken to be dependent on $p_{\eta,i}$ in order to correctly describe the stochastic dynamics of a non-linear oscillator. The explicit form of $D_{\eta,i}(p)$ is reported in Appendix~\ref{app:method}.

In order to integrate the resulting stochastic differential equation we use the Euler-Heun algorithm as implemented in the DifferentialEquations.jl package \cite{rackauckasDifferentialEquationsJlPerformant2017}. As initial conditions we take the phase $\phi_{\eta,i}$ to be uniformly and randomly distributed between $0$ and $2\pi$ and the power $p_{\eta,i}$ to be distributed according to the equilibrium Boltzmann distribution 
\begin{equation}
	P_{eq} \propto \exp \left[ - \frac{2\lambda_{\eta,i}}{k_B T} p_{\eta,i}\right],
\end{equation}
where $\lambda_{\eta,i}$ is a scale factor relating the dimensionless oscillator power $p$ and the oscillator energy.

Since this system is inherently stochastic, both from the initial conditions and the thermal fluctuations, we collect statistics by running every configuration of parameters $\NN$ times. The main observable we are interested in is the number of lasing modes, where a lasing mode is defined as any mode that has power $p_{\eta,i} \geq \epsilon p_0$, where $p_0$ is the steady-state power of a single oscillator \cite{slavinNonlinearAutoOscillatorTheory2009}. We let the system run for a time $t_{\mathrm{end}}$ and choose $\epsilon=0.9$ to account for the fluctuations around the equilibrium power of a STO. 
Our stochastic simulations may not capture all possible processes, such as rare low probability events, in a single run. However, by running multiple 'trajectories' we can gather statistics and gain insight into the behavior of the system. This is also true of experimental runs, where the number of lasing modes will vary in any one realization of the experiment. We will discuss this problem in greater detail in Sec.~\ref{sec:nucleation}, where we also show how the experimental observation times should be chosen.

\section{Results}
\label{sec:results}
\begin{figure*}
	\includegraphics{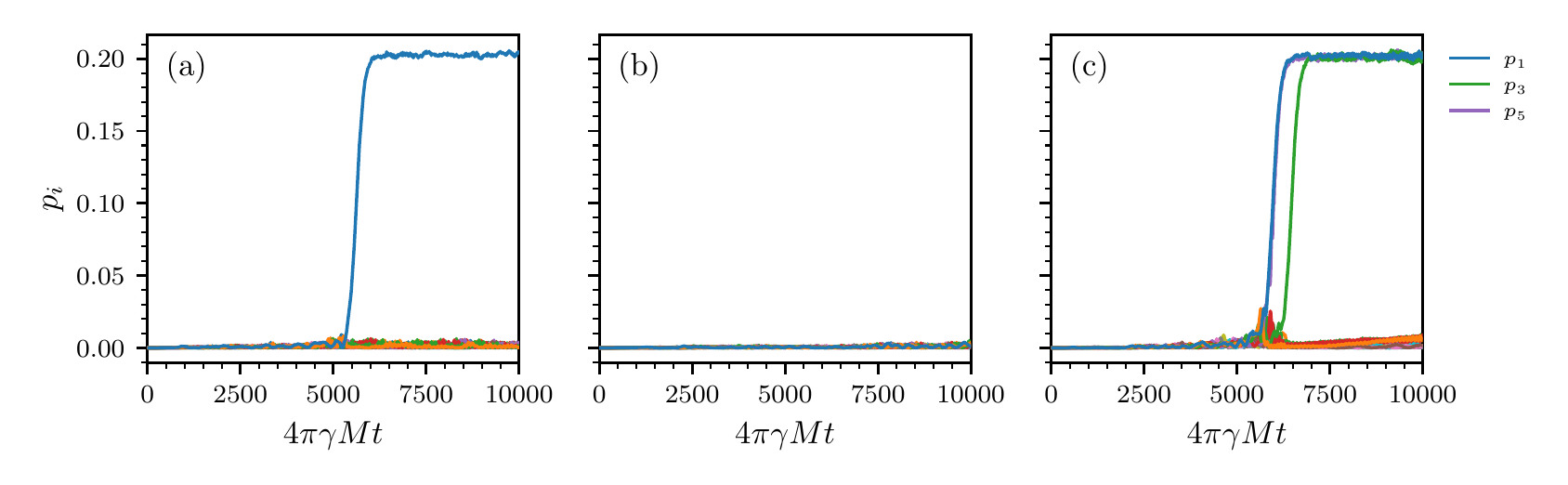}
	\caption{Three types of behavior of the system, with $\tilde{J}=-0.025/4\pi\gamma M$, $J/\tilde{J}=0.5$, $k_B T/\lambda=10^{-5}$ and $G=\dip=0$. (a) Only the edge mode starts lasing, while all the bulk modes are suppressed. (b)  No modes start lasing within the specified time frame. (c) Both the edge modes and bulk modes start lasing. The simulation is run for $4\pi \gamma M t=10^5$, which for a typical STO with $4\pi\gamma M=\SI{10}{GHz}$ is equal to $\SI{10}{\mu s}$. \label{fig:typical}}
\end{figure*}

In this section, we present the results from the simulations described in Sec.~\ref{sec:simulations}. Unless stated otherwise, we set $\alpha=10^{-2}$, $\omega/4\pi\gamma M=0.5$, $J_{s,A}=2\alpha\omega$, $J_{s,B}=0$ and $G=\dip=0$, such that we are in the \PT-symmetric regime and work with an array of $N=10$ unit cells. We run the simulations of Eqs.~(\ref{eq:single-sto}-\ref{eq:sto-coupled-B}) for $4\pi\gamma M t=10^5$, which for a typical STO with $4\pi\gamma M=\SI{10}{GHz}$ corresponds to $\SI{10}{\mu s}$ and collect statistics over $\NN=100$ runs. We are interested in two main observables: whether the edge mode starts lasing, and how many bulk modes also start lasing. It is worth noting that since the B-sites dynamics are suppressed because they are not directly driven, we only have $N$ possible lasing modes. We choose $k_B T/\lambda$ in the range $10^{-6}$ to $10^{-4}$, with the latter corresponding to room temperature for a typical STO \cite{slavinNonlinearAutoOscillatorTheory2009}. 

In Fig.~\ref{fig:typical} we show three typical examples of the system dynamics. The system is initially in thermal equilibrium, and at $t=0$ the spin-torque current is turned on for all A-sites. Fig.~\hyperref[fig:typical]{\ref*{fig:typical}a} show the case where after some time the left-most mode starts lasing at the steady-state power for a single oscillator, whilst the dynamics of the bulk modes are suppressed. Alternatively, no modes can start lasing at all (Fig.~\hyperref[fig:typical]{\ref*{fig:typical}b}), which we will discuss further in Sec.~\ref{sec:nucleation}. We also observed the lasing of bulk modes together with the edge mode, as shown in Fig.~\hyperref[fig:typical]{\ref*{fig:typical}c}. In this specific example the bulk mode starts lasing shortly after the edge mode. We have not investigated this timing further, but it seems likely that a lasing edge mode could also excite bulk modes close to the edge. Moreover, we also observed cases where bulk modes start lasing later in time, seemingly independent of the lasing of the edge mode.

Since all three cases are possible, we further explore the parameter space, and focus on the amount of lasing modes as an observable. We note that in all of the cases discussed here we never observed a lasing bulk mode without a lasing edge mode. This is a direct result of the topology of the array. 

\begin{figure}
	\centering
	\includegraphics{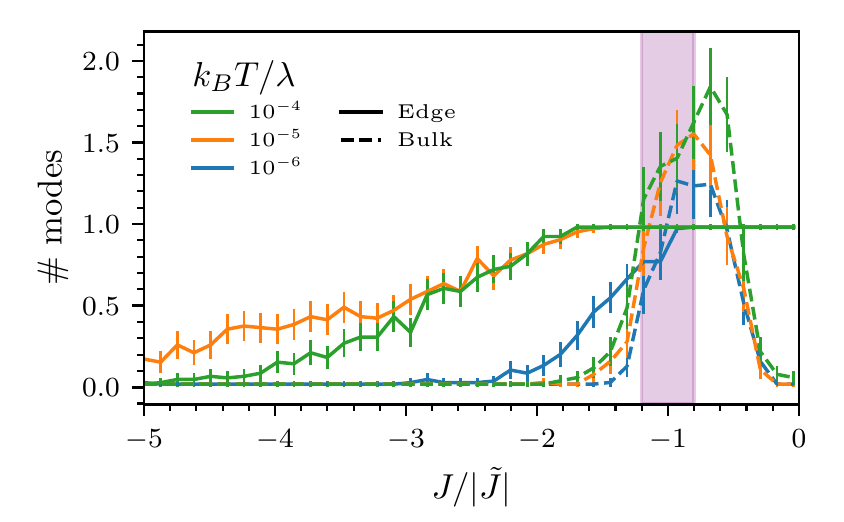}
	\caption{Average number of lasing bulk (dashed line) and edge (solid line) modes as a function of $|J|$, for different temperatures. The \PT-broken regime is indicated by the shaded region. Note that these simulations were only run for $4\pi\gamma Mt=10^5$ and this is therefore only a snapshot of the number of lasing modes. The error bars show the 95\% confidence interval.  \label{fig:EP-array}}
\end{figure}

As discussed before, the system hosts a lasing topological edge state for $-1 < J/|\tilde{J}| < 1$. We thus show the average number of lasing bulk (dashed line) and edge (solid line) modes as a function of $J/|\tilde{J}|$ for different temperatures in Fig.~\ref{fig:EP-array}. The transition from the topological to the trivial regime at $ J/|\tilde{J}| = -1$ is affected by the temperature. For low temperatures the transition is sharper than for high temperatures. However, at high temperatures the system still exhibits signs of a non-Hermitian topological insulator (suppressing of the bulk modes, with only a single lasing edge mode), even if the system is in the trivial phase, i.e. if $|J|>|\tilde{J}|$.

As was previously discussed, the system has broken \PT-symmetry when $|\tilde{J}-\alpha\omega| < |J| < |\tilde{J}+\alpha\omega|$, where multiple modes will start to lase. In Fig.~\ref{fig:EP-array} the \PT-broken regime is indicated by the shaded area and it is easy to see that more bulk  modes start lasing. For higher temperatures more bulk modes will start to lase. We remark here that these simulations were only run for a fixed time $\tend$, and therefore not all modes might have started lasing yet. We return to this issue in Sec.~\ref{sec:nucleation}. 

The number of lasing bulk modes also increases as a function of temperature. Instead of a sharp transition at the exceptional point (EP), there is a transitional regime, due to the non-linear and stochastic effects. This indicates that, depending on the operating temperature, it is necessary to stay further away from the EP than one might initially expect. Specifically for this parameter set it would mean choosing $|J|\ll |\tilde{J}+\alpha\omega|$, such that there are no unwanted bulk contributions from the \PT-broken regime. The fact that the \PT-broken regime extends further then expected might also have implications for applications using the exceptional point, such as enhanced sensing \cite{chenExceptionalPointsEnhance2017,miriExceptionalPointsOptics2019} and encircling the exceptional point \cite{dembowskiEncirclingExceptionalPoint2004}.

\begin{figure*}
	\includegraphics{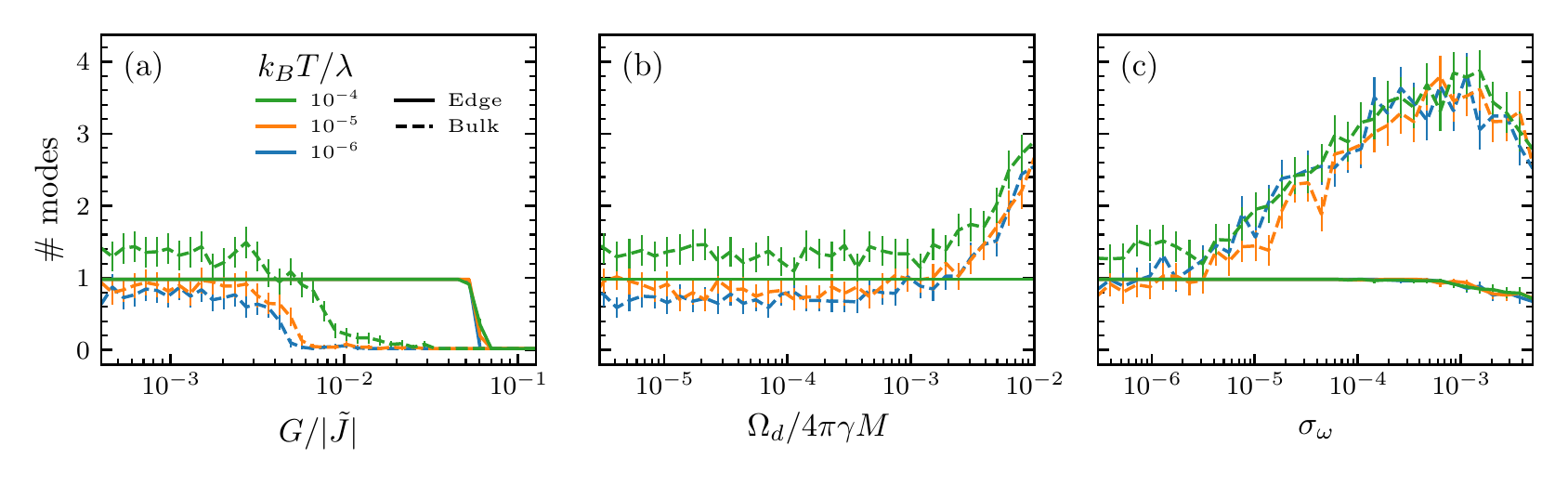}
	\caption{Average number of lasing bulk (dashed line) and edge (solid line) modes for different perturbations that will always be present in any realistic system. The error bars show the 95\% confidence interval. Here $\tilde{J}=-0.025/4\pi\gamma M$, $J/\tilde{J}=0.5$. (a) As a function of the strength of the dissipative coupling $G=G_i=\tilde{G}_i$, which is mediated by the spacer layers between the STOs. (b) As a function of the dipole-dipole coupling strength $\dip / 4\pi \gamma M$, which can be tuned by the spacing between the STOs (c) As function of the variation of the individual FMR-frequency $\omega_{\eta,i}$, chosen from a normal distribution with mean $\omega/4\pi\gamma M=0.5$ and standard deviation $\sigma_\omega$. \label{fig:disturbances}
	}
\end{figure*}

\subsection{Sensitivity to perturbations}
We now consider three main perturbations present in any experimental realization of the system: (1) the dissipative coupling modulated by spin-waves traveling in the metallic spacer layer, (2) the dipole-dipole coupling between the macrospins of the STOs and (3) variations in the parameters of the individual STOs. 

We first consider the inter (intra)-layer dissipative coupling $G$ ($\tilde{G}$), induced by the metallic spacer layer. This coupling is known to synchronize STOs \cite{slavinTheoryMutualPhase2006}. Since it is modulated by the metallic spacer layer it can be tuned to an extent, e.g. by choosing a spacer layer with a certain spin relaxation. It is also possible to choose a STO nano-pillar geometry in which the metallic spacer layer does not extend in-between the STOs and therefore no spin waves can propagate, thus suppressing the dissipative coupling \cite{slavinNonlinearAutoOscillatorTheory2009}. 

We vary the dissipative coupling $G=G_i=\tilde{G}_i$, the results of which are shown in Fig.~\hyperref[fig:disturbances]{\ref*{fig:disturbances}a}. We note here that even though the dissipative coupling breaks the \PT-symmetry, the edge modes are still protected by CI-symmetry. When $G/|\tilde{J}|> 10^{-3}$, first the bulk modes are suppressed. This can be attributed to the increased overall dissipation in the system, suppressing the bulk excitations. When $G/|\tilde{J}|>0.05 $ the edge modes are also suppressed, which again can be attributed to the increased overall dissipation. This result thus suggests that it is desirable to design a system where the dissipative coupling is weak, such as by using the nano-pillar geometry. However, a small dissipative coupling might be beneficial, effectively suppressing the bulk excitations, while allowing the edge modes to lase.

Next, we discuss the dipole-dipole coupling, which is present in any magnetic system. It has also been known to synchronize the precession in STOs \cite{slavinTheoryMutualPhase2006}. The dipolar coupling strength can be controlled by the spacing between the STOs.
We vary the dipolar coupling strength $\dip/4\pi\gamma M$ for different temperatures, as shown in Fig.~\hyperref[fig:disturbances]{\ref*{fig:disturbances}b}. For small $\dip$ we see no changes, and only for $\dip/4\pi\gamma M > 10^{-3}$ the bulk modes will start to lase. For STOs of typical dimensions $10\times10\times\SI{10}{nm}$ and a separation distance \SI{10}{nm}, $\dip/4\pi\gamma M\approx10^{-3}$. Our results indicate a lower bound on the spacing between STOs in order to avoid activation of the bulk lasing modes due to the dipole-dipole interaction. 

In any experimental setup there will be small variations between the individual STOs. Since the system considered here is only -symmetric if the dissipation is balanced with the driving, even small variations in any parts of the STO array involved in the driving and dissipation processes will break the \PT-symmetry. Since the topological classification of this system is based on the \PT-symmetry, it is useful to consider the effect of breaking this symmetry.

In order to model spatial disorder we consider an array with small variations in the individual frequencies $\omega_i$, by assuming they are normally distributed with standard deviation $\sigma_{\omega}$ and mean $\omega$.  If the local spin-transfer torque is kept constant at $J_s=2\alpha\omega$ throughout the array, the system is no longer \PT-symmetric. This spin-injection model corresponds to setup in which a single current source, rather than individual ones, are used to inject spin angular momentum into the STOs. The number of lasing modes is shown in Fig.~\hyperref[fig:disturbances]{\ref*{fig:disturbances}c}, where it is clear that as the variance is increased, more bulk modes start lasing. The disorder we introduce breaks the \PT-symmetry, and the edge states are no longer topologically protected. The bulk modes are therefore no longer suppressed and can start lasing. We have also modeled the case where $J_{s,i}=2\alpha\omega_i$, i.e. where the STOs are individually driven. This did not affect the results, indicating that it is not just the \PT-symmetry within one unit cell, but rather the \PT-symmetry of the complete array that protects the edge states.
The robustness of non-Hermitian topological states against disorder is still poorly understood. From the linear dynamics we know that if a variance $\sigma_{\omega}$ is introduced the bulk modes also gain a non-zero imaginary component and will therefore start lasing or be suppressed. This is thus in essence not a non-linear or stochastic effect, in contrast to the other effects discussed previously.

We thus conclude that the non-Hermitian SSH chain can be experimentally realized using STOs, but care needs to be taken to control the dissipative coupling, dipolar interactions and the variations between STOs. 

\subsection{Nucleation}
\label{sec:nucleation}
The processes that we consider in this work are fully stochastic, and so is the lasing of the edge mode: there is a finite probability that the edge mode will start lasing. 
It is therefore possible that even in the topological phase, the edge mode might only start lasing at time scales longer than the experimental observation time. Motivated by this consideration, we thus investigate here the nucleation times of the lasing edge modes. 

\begin{figure}
	\includegraphics{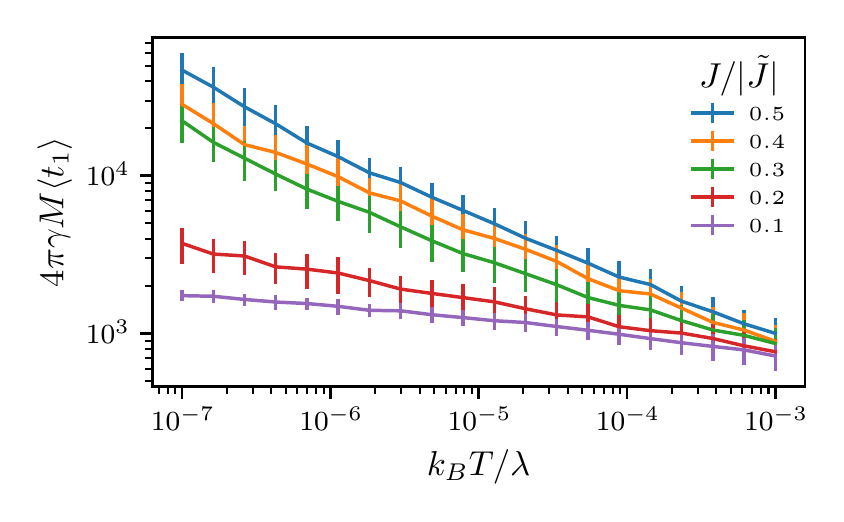}
	\caption{The average time for the edge mode to start lasing, for different ratios $J/|\tilde{J}|$. These simulations are all in the unbroken \PT-regime. For a typical FMR frequency $4\pi\gamma M=\SI{10}{GHz}$ these times are in the order of \SIrange{.01}{1}{ms}. The error bars show one standard deviation. \label{fig:first-laser}}
\end{figure}

As before, we prepare our system in thermal equilibrium, and turn on the spin current at $t=0$. We only consider the RKKY-type coupling, and set $G=\tilde{G}=\dip=0$. In Fig.~\ref{fig:first-laser} we show the temperature dependence of the time at which the edge mode starts lasing, for different ratios $J/|\tilde{J}|$. We have chosen the ratio $J/|\tilde{J}|$ such that we are in the unbroken \PT~regime and we only expect the edge state to start lasing.

It is clear that the time until the edge mode starts lasing follows an exponential distribution as a function of temperature. Moreover, for lower intra-cell coupling $J$, the average time decreases. This observation can be explained in terms of overcoming an energy barrier. The probability of the edge mode to start lasing is
\begin{equation}
	P_1 \propto \exp \left[-\frac{\Delta E}{k_B T}\right],
\end{equation}
where $\Delta E$ is the energy difference between the state with no modes lasing and the state with a lasing edge mode, which is directly related to the coupling strength $J$.

The nucleation problem also illustrates that the finite runtimes inherent with numerical simulations might not be representative of experiments. Since the nucleation is a stochastic process, it is possible that for longer runtimes, such as seconds, all modes will start lasing. In order to have robust isolated edge modes, these devices might therefore be limited to shorter runtimes (on the order of \si{\micro s}), especially at higher temperatures. The opposite problem of course also exist, where one has to wait a long time for the edge mode to start lasing. The nucleation time can however be tuned with the exchange coupling, as is shown in Fig.~\ref{fig:first-laser}. The inherent stochastic nature of this device could make it useful for stochastic computing, which relies on systems with inherent randomness \cite{vincentSpinTransferTorqueMagnetic2015,grollierNeuromorphicSpintronics2020}.

\section{Comparison to photonic systems}
\label{sec:photonic}
The array of STOs considered here shares many similarities with the photonic microring resonators experimentally realized in Ref.~\cite{partoEdgeModeLasing1D2018}. The linear dynamics of the magnetic excitations considered in this model are closely related to the dynamics of the photonic excitations in the microring resonators. Both models realize the non-Hermitian SSH model as described by Eq.~(\ref{eq:ham}). Moreover, their non-linear dynamics are similar, as we show next.

The dynamics of the microring resonator array are, up to second order in the (normalized) electric modal field amplitudes $a_n^\eta$, described by \cite{hassanNonlinearReversalMathcalPT2015}
\begin{equation}
	\begin{aligned}
	\partial_t a_n^A &= -i\Omega a_n^A -\gamma -  \sigma_A \left( 1 - |a_n^A|^2 \right) \\
	&\qquad + i \kappa_1 a_n^B + i \kappa_2 a_{n-1}^B, \\ 
	\partial_t a_n^B &= -i\Omega a_n^B -\gamma -  \sigma_B \left( 1 - |a_n^B|^2 \right) \\
	&\qquad+ i \kappa_1 a_n^A + i \kappa_2 a_{n+1}^A,
	\end{aligned}
	\label{eq:photonics}
\end{equation}
where $\Omega$ is the lasing mode frequency, $\gamma$ is the mode loss, $\sigma_\eta$ is the mode gain and $\kappa_{1,2}$ are the coupling constants.
Thus, we can draw a direct analogue between the two systems, by identifying $\gamma$ as the constant part of the Gilbert damping ($\gamma=\alpha\omega$), the lasing mode frequency $\Omega$ as the precession frequency ($\Omega=\omega$) and the mode gain as the spin current contribution ($\sigma_{\eta}=J_{s,\eta}$). One distinction is the linear coupling $\kappa_{1,2}$, whereas the RKKY coupling between the STOs is non-linear. However, we note that up to first order in $p_i$ the RKKY coupling is linear as well.  In order to fully describe the dynamics of the microring resonators the carrier density also has to be taken into account. However, for timescales longer than the time-response of the laser, typically a few nanoseconds \cite{agrawalSemiconductorLasers1995}, the carrier dynamics can be disregarded and one obtains Eq.~(\ref{eq:photonics}). 

Most importantly, both systems have saturated gain. This feature is inherent to many driven non-Hermitian system, if the driving is limited in some way.
We do note that the non-linear contributions to the precession and Gilbert damping that are present in STO systems are not present in photonics. The phase diagram, as presented in Fig.~\ref{fig:EP-array}, might thus be different for the photonic system.

In this work we have considered three perturbations: dissipative couplings, dipole-dipole interactions and variations in the FMR-frequency, as shown in Fig.~\ref{fig:disturbances}. Dissipative coupling can also be present in photonic systems, but when employing evanescent coupling is usually negligible \cite{dingModeDiscriminationDissipatively2019}, indicating that it is less relevant for photonic implementations of the non-Hermitian SSH chain. The long-range dipole-dipole interaction has no photonic analogue, but variations in the parameters are inevitably present in photonic systems.

Non-linear effects are a common feature of many experimental realizations of non-Hermitian systems. Our results are therefore applicable beyond STO arrays to other non-Hermitian topological phases.

\section{Conclusion and discussion}
\label{sec:conclusion}
In conclusion, we have shown that the lasing topological edge states can be successfully accessed using STOs in a realistic non-Hermitian SSH array. We have considered both non-linear and stochastic dynamics, to determine if an experimental implementation of this model is feasible. 
Firstly, we found that the lasing edge mode is robust in the presence of a wide range of temperatures and perturbations, provided that we are in the topological regime. This occurs despite various perturbations breaking the PT-symmetry of the original Hamiltonian such that topological protection is not guaranteed. Our result is important for ensuring that the lasing edge mode can be probed in an experiment. 
Secondly, we found that even though the system is topological in the linear regime, in which no bulk modes should start to lase, non-linear and stochastic effects can still access these bulk modes, reducing the usefulness of the topologically protected edge modes. We have explored the transition between the \PT-unbroken and \PT-broken regimes, which is not a sharp transition. Instead, we find a regime around the exceptional point where more bulk modes will start to lase. 

Moreover, we have considered three kinds of perturbations that can naturally be present in this array and have shown in which regime the topology of the system is unaffected. We hope that these results can be used to guide future experiments. We find that at a given temperature and for equal strength of the perturbative term, the perturbation that mostly affects the dynamics of bulk modes is the variation in the parameters of the individual STOs. This might complicate the experimental realization of the STO array, and will be an inherent complication in any physical realization of a non-Hermitian SSH model, where variations are more likely. Finally, we have shown that there is a finite nucleation time, after which the edge state will start lasing. 

Interactions between STOs as considered here are not easily tunable, since they depend on the spacing between the STOs \cite{slavinTheoryMutualPhase2006}. Moreover, STOs cannot be brought arbitrarily close together because of Joule heating  \cite{slavinSpintorqueOscillatorsGet2009}. To circumvent this problem, one could couple the STOs using strip-line antennas above each STO, controlled by the microwave current generated at the adjacent STO \cite{wittrockExceptionalPointsControlling2021}. This more direct method has the advantage that the signal could be electrically amplified, thus allowing control over the coupling strength. Even though this coupling has a different physical origin, the conclusions as presented in this work will still hold.

\begin{acknowledgments}
	R.D. is member of the D-ITP consortium, a program of the Dutch Organization for Scientific Research (NWO) that is funded by the Dutch Ministry of Education, Culture and Science (OCW). This project has received funding from the European Research Council (ERC) under the European Union’s Horizon 2020 research and innovation programme (grant agreement No. 725509). This work is part of the research programme of the Foundation for Fundamental Research on Matter (FOM), which is part of the Netherlands Organization for Scientific Research (NWO). This work is part of the Fluid Spintronics research programme with project number 182.069,
	which is financed by the Dutch Research Council (NWO) B.F. acknowledges support of the National Science Foundation under Grant No. NSF DMR-2144086.
\end{acknowledgments}

\appendix

\section{Numerical implementation}
\label{app:method}
A single STO $\eta$ in unit cell $i$ can be parametrized with the complex amplitude $c_{\eta,i}(t)$ \cite{slavinNonlinearAutoOscillatorTheory2009}
\begin{equation}
	\mathbf{m}_{\eta,i}=M_{\eta,i} \begin{pmatrix}
		\sqrt{1-p_{\eta,i}}\left(c_{\eta,i} + c_{\eta,i}^*\right) \\
		\sqrt{1-p_{\eta,i}}\left(ic_{\eta,i} - ic_{\eta,i}^*\right) \\
		1-2p_{\eta,i}
	\end{pmatrix},
\end{equation}
where $p_{\eta,i}=|c_{\eta,i}|^2$ and $M_{\eta,i}$ is the magnetization length. The Langevin equation of motion then becomes
\begin{multline}
	\frac{\dd{c}_{\eta,i}}{\dd{t}} + i\omega_{\eta,i}\left(p_{\eta,i}\right)c_{\eta,i} + \Gamma_{\eta,i}^+\left(p_{\eta,i}\right)c_{\eta,i}  \\
	-\Gamma_{\eta,i}^-\left(p_{\eta,i}\right)c_{\eta,i} = f_{\eta,i}(t),
	\label{eq:langevin}
\end{multline}
where 
\begin{align}
	\omega_{\eta,i}\left(p\right) &= \omega + 2p,\\ 
	\Gamma_{\eta,i}^+\left(p\right) &= \omega_{\eta,i}\alpha_{\eta,i} \left(1+\left(2/\omega_{\eta,i} - 1\right)p\right), \\
	\Gamma_{\eta,i}^-\left(p\right) &= J_{s;\eta,i} \left(1-p\right),
\end{align}
where the superscript $+$ ($-$) indicates loss (gain) and $f_{\eta,i}(t)$ is a complex field, representing the thermal fluctuations. The stochastic field is a phenomenological description of all thermal processes and chosen to have zero mean and a second-order correlator
\begin{equation}
	\begin{gathered}
		\langle f_{\eta,i}(t)f_{\eta',j}(t') \rangle = 0,\\ 
		\langle f_{\eta,i}(t) f_{\eta',j}^*(t') \rangle = 2\delta_{i,j}\delta_{\eta,\eta'} D_{\eta,i}\left(p_{\eta,i}\right) \delta\left(t-t'\right),
	\end{gathered}
\end{equation}
The diffusion coefficient $D_{\eta,i}(p)$ has to be taken dependent on $p$ such that the system tends to thermal equilibrium. This is done by deriving the Fokker-Planck equation from the Langevin Eq.~(\ref{eq:langevin}) and finding a physically-consistent solution \cite{tiberkevichMicrowavePowerGenerated2007}. For a single STO this is 
\begin{equation}
	D_{\eta,i}(p)=\Gamma^+_{\eta,i}(p) \frac{k_B T}{\lambda_{\eta,i} \omega_{\eta,i}(p)},
\end{equation}
where $\lambda_{\eta,i}$ is a scale factor relating the dimensionless oscillator power $p$ and the oscillator energy, which is $\lambda_{\eta,i}=V_{\mathrm{eff{\eta,i}}}M_{\eta,i}/\gamma_{\eta,i}$ for our choice of parametrization.

As initial conditions we choose $c_{\eta,i}=\sqrt{p_0}e^{i\phi}$, where $\phi$ is randomly chosen between $0$ and $2\pi$ and $p_0$ is drawn from a thermal equilibrium distribution $P_{\mathrm{eq}}$ for an ensemble of isolated STOs
\begin{equation}
	P_{\mathrm{eq}} \propto \exp\left[-\frac{2\lambda_{\eta,i}}{k_B T} p_{\eta,i}\right].
\end{equation}

For the array as considered in the main text, the equation of motion is given by 
\begin{align}
	\partial_t c_{A,i} &= \fsto(c_{A,i}) c_{A,i} \nonumber\\
	&\quad + J_i \fj(c_{A,i},c_{B,i}) + \tilde{J}_{i-1}\fj(c_{A,i},c_{i-1,B})\nonumber \\
	&\quad + G_i \fg(c_{A,i},c_{B,i}) + \tilde{G}_{i-1}\fg(c_{A,i},c_{i-1,B}) \nonumber\\
	&\quad + \dip \sideset{}{'}\sum_{\eta,j} \frac{\fdip(c_{A,i}, c_{j,\eta})}{r_{Ai,\eta j}^3}, \\
	\partial_t c_{B,i} &= \fsto(c_{B,i}) c_{B,i} \nonumber\\
	&\quad + J_i \fj(c_{B,i},c_{A,i}) + \tilde{J}_{i}\fj(c_{B,i},c_{i+1,A})\nonumber \\
	&\quad + G_i \fg(c_{B,i},c_{A,i}) + \tilde{G}_{i-1}\fg(c_{B,i},c_{i+1,A}),\nonumber\\
	&\quad + \dip \sideset{}{'}\sum_{\eta,j} \frac{\fdip(c_{B,i}, c_{j,\eta}) }{r_{Bi,\eta j}^3},
\end{align}
where $\fj(c_{\eta,i},c_{j,\eta'})$,  $\fg(c_{\eta,i},c_{j,\eta'})$ and $\fdip(c_{B,i}, c_{j,\eta})$ are the RRKY, dissipative and dipolar couplings between the $\mathrm{STO}_{\eta i}$ and $\mathrm{STO}_{\eta' j}$. These are given by
\begin{widetext}

\begin{align}
	\fsto(c_{\eta,i}) &= -i (\omega_{\eta,i} + 2p_{\eta,i}) - \alpha_{\eta,i} \omega_{\eta,i} (1 + (2/\omega_{\eta,i} - 1)p_{\eta,i}) + J_{s,\eta,i} (1-p_{\eta,i}), \\
	\fj(c_{\eta,i},c_{j,\eta'}) &= \frac{i}{2\sqrt{1-p_{\eta,i}}} \Bigr(  c_{j,\eta'} (2-3p_{\eta,i}) \sqrt{1-p_{j,\eta'}} \nonumber\\
	&\qquad+ c_{\eta,i}c^*_{j,\eta'} (4c_{j,\eta'} \sqrt{1-p_{\eta,i}} - c_{\eta,i} \sqrt{1-p_{j,\eta'}}) -2c_{\eta,i} \sqrt{1-p_{\eta,i}} \Bigr),\\
	\fg(c_{\eta,i},c_{j,\eta'}) &= \frac{1}{1-p_{\eta,i}} ( c_{j,\eta'} \omega_{j,\eta'} \left( -\sqrt{1-p_{j,\eta'}}\sqrt{1-p_{\eta,i}} + c_{j,\eta'}^*( 2c_{j,\eta'} \sqrt{1-p_{j,\eta'}}\sqrt{1-p_{\eta,i}} -c_{\eta,i}) \right) \nonumber\\
	&\qquad- 2c_{j,\eta'}p_{j,\eta'} (\sqrt{1-p_{j,\eta'}}\sqrt{1-p_{\eta,i}} + c_{\eta,i} c_{j,\eta'}^*) 
	 ),\\
	\fdip(c_{\eta,i},c_{j,\eta'}) &= \frac{i}{\sqrt{1-p_{\eta,i}}} (  c_{\eta,i} \sqrt{1-p_{\eta,i}} + 2c_{\eta,i}^2 c_{j,\eta'}^* \sqrt{1-p_{j,\eta'}} -2c_{\eta,i} p_{j,\eta'} \sqrt{1-p_{\eta,i}} \nonumber\\
	&\qquad- \sqrt{1-p_{j,\eta'}} \left( 1 - 3c_{\eta,i} \Re[c_{\eta,i}]\right) (c_{j,\eta'} - 3\Re[c_{j,\eta'}]) ).
\end{align}

\section{Hamiltonian}
\label{app:ham}

The full Hamiltonian is given by $H=\sum_i H_i + H_i^G + H^{\mathrm{dip}}_i$, where $H_i$ is given in Eq.~(\ref{eq:ham}) and
\begin{align}
	H_i^G &= iG\omega\left[ a_i^\dagger b_i + \hc \right] + i\tilde{G}\omega\left[ a_i^\dagger b_{i-1} + \hc \right], \\
	H_i^{\mathrm{dip}} &=  \dip\sum_{{\eta}} \left[{\eta}_i^\dagger{\eta}_i + \sideset{}{'}\sum_{j,\eta'} \frac{{\eta}_i^\dagger {\eta'}_j + 3{\eta}_i^\dagger {\eta'}_j^\dagger + 3{\eta}_i{\eta'}_j}{2r_{{\eta i}, \eta'j}^3}  \right],
\end{align}
with $\eta,\eta'=a,b$ and the $\sum\nolimits'$ indicates that the sum excludes self-interactions (i.e. $i=j,\nu=\eta$). The boundary conditions are then given by Eq.~(\ref{eq:ham-bc}) and
\begin{equation}
	H_j^G = iG\omega\left[ a_j^\dagger b_j + \hc \right] + i\tilde{G}\omega \left[ b_j^\dagger a_l + \hc \right].
\end{equation}

\end{widetext}


\begin{thebibliography}{58}%
	\makeatletter
	\providecommand \@ifxundefined [1]{%
		\@ifx{#1\undefined}
	}%
	\providecommand \@ifnum [1]{%
		\ifnum #1\expandafter \@firstoftwo
		\else \expandafter \@secondoftwo
		\fi
	}%
	\providecommand \@ifx [1]{%
		\ifx #1\expandafter \@firstoftwo
		\else \expandafter \@secondoftwo
		\fi
	}%
	\providecommand \natexlab [1]{#1}%
	\providecommand \enquote  [1]{``#1''}%
	\providecommand \bibnamefont  [1]{#1}%
	\providecommand \bibfnamefont [1]{#1}%
	\providecommand \citenamefont [1]{#1}%
	\providecommand \href@noop [0]{\@secondoftwo}%
	\providecommand \href [0]{\begingroup \@sanitize@url \@href}%
	\providecommand \@href[1]{\@@startlink{#1}\@@href}%
	\providecommand \@@href[1]{\endgroup#1\@@endlink}%
	\providecommand \@sanitize@url [0]{\catcode `\\12\catcode `\$12\catcode
		`\&12\catcode `\#12\catcode `\^12\catcode `\_12\catcode `\%12\relax}%
	\providecommand \@@startlink[1]{}%
	\providecommand \@@endlink[0]{}%
	\providecommand \url  [0]{\begingroup\@sanitize@url \@url }%
	\providecommand \@url [1]{\endgroup\@href {#1}{\urlprefix }}%
	\providecommand \urlprefix  [0]{URL }%
	\providecommand \Eprint [0]{\href }%
	\providecommand \doibase [0]{https://doi.org/}%
	\providecommand \selectlanguage [0]{\@gobble}%
	\providecommand \bibinfo  [0]{\@secondoftwo}%
	\providecommand \bibfield  [0]{\@secondoftwo}%
	\providecommand \translation [1]{[#1]}%
	\providecommand \BibitemOpen [0]{}%
	\providecommand \bibitemStop [0]{}%
	\providecommand \bibitemNoStop [0]{.\EOS\space}%
	\providecommand \EOS [0]{\spacefactor3000\relax}%
	\providecommand \BibitemShut  [1]{\csname bibitem#1\endcsname}%
	\let\auto@bib@innerbib\@empty
	\bibitem [{\citenamefont {Hasan}\ and\ \citenamefont
		{Kane}(2010)}]{hasanColloquiumTopologicalInsulators2010a}%
	\BibitemOpen
	\bibfield  {author} {\bibinfo {author} {\bibfnamefont {M.~Z.}\ \bibnamefont
			{Hasan}}\ and\ \bibinfo {author} {\bibfnamefont {C.~L.}\ \bibnamefont
			{Kane}},\ }\bibfield  {title} {\bibinfo {title} {Colloquium: {{Topological}}
			insulators},\ }\href {https://doi.org/10.1103/RevModPhys.82.3045} {\bibfield
		{journal} {\bibinfo  {journal} {Reviews of Modern Physics}\ }\textbf
		{\bibinfo {volume} {82}},\ \bibinfo {pages} {3045} (\bibinfo {year}
		{2010})},\ \bibinfo {note} {overzicht van topological insulators}\BibitemShut
	{NoStop}%
	\bibitem [{\citenamefont {Bender}\ \emph {et~al.}(2002)\citenamefont {Bender},
		\citenamefont {Brody},\ and\ \citenamefont
		{Jones}}]{benderComplexExtensionQuantum2002}%
	\BibitemOpen
	\bibfield  {author} {\bibinfo {author} {\bibfnamefont {C.~M.}\ \bibnamefont
			{Bender}}, \bibinfo {author} {\bibfnamefont {D.~C.}\ \bibnamefont {Brody}},\
		and\ \bibinfo {author} {\bibfnamefont {H.~F.}\ \bibnamefont {Jones}},\
	}\bibfield  {title} {\bibinfo {title} {Complex {{Extension}} of {{Quantum
					Mechanics}}},\ }\href {https://doi.org/10.1103/PhysRevLett.89.270401}
	{\bibfield  {journal} {\bibinfo  {journal} {Physical Review Letters}\
		}\textbf {\bibinfo {volume} {89}},\ \bibinfo {pages} {270401} (\bibinfo
		{year} {2002})}\BibitemShut {NoStop}%
	\bibitem [{\citenamefont {Bender}\ and\ \citenamefont
		{Boettcher}(1998)}]{benderRealSpectraNonHermitian1998}%
	\BibitemOpen
	\bibfield  {author} {\bibinfo {author} {\bibfnamefont {C.~M.}\ \bibnamefont
			{Bender}}\ and\ \bibinfo {author} {\bibfnamefont {S.}~\bibnamefont
			{Boettcher}},\ }\bibfield  {title} {\bibinfo {title} {Real {{Spectra}} in
			{{Non-Hermitian Hamiltonians Having}} $\mathcal{PT}$ {{Symmetry}}},\ }\href
	{https://doi.org/10.1103/PhysRevLett.80.5243} {\bibfield  {journal} {\bibinfo
			{journal} {Physical Review Letters}\ }\textbf {\bibinfo {volume} {80}},\
		\bibinfo {pages} {5243} (\bibinfo {year} {1998})}\BibitemShut {NoStop}%
	\bibitem [{\citenamefont {Bergholtz}\ \emph {et~al.}(2021)\citenamefont
		{Bergholtz}, \citenamefont {Budich},\ and\ \citenamefont
		{Kunst}}]{bergholtzExceptionalTopologyNonHermitian2021}%
	\BibitemOpen
	\bibfield  {author} {\bibinfo {author} {\bibfnamefont {E.~J.}\ \bibnamefont
			{Bergholtz}}, \bibinfo {author} {\bibfnamefont {J.~C.}\ \bibnamefont
			{Budich}},\ and\ \bibinfo {author} {\bibfnamefont {F.~K.}\ \bibnamefont
			{Kunst}},\ }\bibfield  {title} {\bibinfo {title} {Exceptional topology of
			non-{{Hermitian}} systems},\ }\href
	{https://doi.org/10.1103/RevModPhys.93.015005} {\bibfield  {journal}
		{\bibinfo  {journal} {Reviews of Modern Physics}\ }\textbf {\bibinfo {volume}
			{93}},\ \bibinfo {pages} {015005} (\bibinfo {year} {2021})}\BibitemShut
	{NoStop}%
	\bibitem [{\citenamefont {Kawabata}\ \emph {et~al.}(2019)\citenamefont
		{Kawabata}, \citenamefont {Shiozaki}, \citenamefont {Ueda},\ and\
		\citenamefont {Sato}}]{kawabataSymmetryTopologyNonHermitian2019}%
	\BibitemOpen
	\bibfield  {author} {\bibinfo {author} {\bibfnamefont {K.}~\bibnamefont
			{Kawabata}}, \bibinfo {author} {\bibfnamefont {K.}~\bibnamefont {Shiozaki}},
		\bibinfo {author} {\bibfnamefont {M.}~\bibnamefont {Ueda}},\ and\ \bibinfo
		{author} {\bibfnamefont {M.}~\bibnamefont {Sato}},\ }\bibfield  {title}
	{\bibinfo {title} {Symmetry and {{Topology}} in {{Non-Hermitian Physics}}},\
	}\href {https://doi.org/10.1103/PhysRevX.9.041015} {\bibfield  {journal}
		{\bibinfo  {journal} {Physical Review X}\ }\textbf {\bibinfo {volume} {9}},\
		\bibinfo {pages} {041015} (\bibinfo {year} {2019})},\ \Eprint
	{https://arxiv.org/abs/1812.09133} {arXiv:1812.09133} \BibitemShut {NoStop}%
	\bibitem [{\citenamefont {Gong}\ \emph {et~al.}(2018)\citenamefont {Gong},
		\citenamefont {Ashida}, \citenamefont {Kawabata}, \citenamefont {Takasan},
		\citenamefont {Higashikawa},\ and\ \citenamefont
		{Ueda}}]{gongTopologicalPhasesNonHermitian2018a}%
	\BibitemOpen
	\bibfield  {author} {\bibinfo {author} {\bibfnamefont {Z.}~\bibnamefont
			{Gong}}, \bibinfo {author} {\bibfnamefont {Y.}~\bibnamefont {Ashida}},
		\bibinfo {author} {\bibfnamefont {K.}~\bibnamefont {Kawabata}}, \bibinfo
		{author} {\bibfnamefont {K.}~\bibnamefont {Takasan}}, \bibinfo {author}
		{\bibfnamefont {S.}~\bibnamefont {Higashikawa}},\ and\ \bibinfo {author}
		{\bibfnamefont {M.}~\bibnamefont {Ueda}},\ }\bibfield  {title} {\bibinfo
		{title} {Topological {{Phases}} of {{Non-Hermitian Systems}}},\ }\href
	{https://doi.org/10.1103/PhysRevX.8.031079} {\bibfield  {journal} {\bibinfo
			{journal} {Physical Review X}\ }\textbf {\bibinfo {volume} {8}},\ \bibinfo
		{pages} {031079} (\bibinfo {year} {2018})}\BibitemShut {NoStop}%
	\bibitem [{\citenamefont {Shen}\ \emph {et~al.}(2018)\citenamefont {Shen},
		\citenamefont {Zhen},\ and\ \citenamefont
		{Fu}}]{shenTopologicalBandTheory2018}%
	\BibitemOpen
	\bibfield  {author} {\bibinfo {author} {\bibfnamefont {H.}~\bibnamefont
			{Shen}}, \bibinfo {author} {\bibfnamefont {B.}~\bibnamefont {Zhen}},\ and\
		\bibinfo {author} {\bibfnamefont {L.}~\bibnamefont {Fu}},\ }\bibfield
	{title} {\bibinfo {title} {Topological {{Band Theory}} for {{Non-Hermitian
					Hamiltonians}}},\ }\href {https://doi.org/10.1103/PhysRevLett.120.146402}
	{\bibfield  {journal} {\bibinfo  {journal} {Physical Review Letters}\
		}\textbf {\bibinfo {volume} {120}},\ \bibinfo {pages} {146402} (\bibinfo
		{year} {2018})}\BibitemShut {NoStop}%
	\bibitem [{\citenamefont {Chiu}\ \emph {et~al.}(2016)\citenamefont {Chiu},
		\citenamefont {Teo}, \citenamefont {Schnyder},\ and\ \citenamefont
		{Ryu}}]{chiuClassificationTopologicalQuantum2016}%
	\BibitemOpen
	\bibfield  {author} {\bibinfo {author} {\bibfnamefont {C.-K.}\ \bibnamefont
			{Chiu}}, \bibinfo {author} {\bibfnamefont {J.~C.~Y.}\ \bibnamefont {Teo}},
		\bibinfo {author} {\bibfnamefont {A.~P.}\ \bibnamefont {Schnyder}},\ and\
		\bibinfo {author} {\bibfnamefont {S.}~\bibnamefont {Ryu}},\ }\bibfield
	{title} {\bibinfo {title} {Classification of topological quantum matter with
			symmetries},\ }\href {https://doi.org/10.1103/RevModPhys.88.035005}
	{\bibfield  {journal} {\bibinfo  {journal} {Reviews of Modern Physics}\
		}\textbf {\bibinfo {volume} {88}},\ \bibinfo {pages} {035005} (\bibinfo
		{year} {2016})}\BibitemShut {NoStop}%
	\bibitem [{\citenamefont {Kawabata}\ \emph {et~al.}(2018)\citenamefont
		{Kawabata}, \citenamefont {Shiozaki},\ and\ \citenamefont
		{Ueda}}]{kawabataAnomalousHelicalEdge2018}%
	\BibitemOpen
	\bibfield  {author} {\bibinfo {author} {\bibfnamefont {K.}~\bibnamefont
			{Kawabata}}, \bibinfo {author} {\bibfnamefont {K.}~\bibnamefont {Shiozaki}},\
		and\ \bibinfo {author} {\bibfnamefont {M.}~\bibnamefont {Ueda}},\ }\bibfield
	{title} {\bibinfo {title} {Anomalous helical edge states in a non-{{Hermitian
					Chern}} insulator},\ }\href {https://doi.org/10.1103/PhysRevB.98.165148}
	{\bibfield  {journal} {\bibinfo  {journal} {Physical Review B}\ }\textbf
		{\bibinfo {volume} {98}},\ \bibinfo {pages} {165148} (\bibinfo {year}
		{2018})},\ \bibinfo {note} {analysis of chern insulator}\BibitemShut
	{NoStop}%
	\bibitem [{\citenamefont {Martinez~Alvarez}\ \emph
		{et~al.}(2018{\natexlab{a}})\citenamefont {Martinez~Alvarez}, \citenamefont
		{Barrios~Vargas},\ and\ \citenamefont
		{Foa~Torres}}]{martinezalvarezNonHermitianRobustEdge2018}%
	\BibitemOpen
	\bibfield  {author} {\bibinfo {author} {\bibfnamefont {V.~M.}\ \bibnamefont
			{Martinez~Alvarez}}, \bibinfo {author} {\bibfnamefont {J.~E.}\ \bibnamefont
			{Barrios~Vargas}},\ and\ \bibinfo {author} {\bibfnamefont {L.~E.~F.}\
			\bibnamefont {Foa~Torres}},\ }\bibfield  {title} {\bibinfo {title}
		{Non-{{Hermitian}} robust edge states in one dimension: {{Anomalous}}
			localization and eigenspace condensation at exceptional points},\ }\href
	{https://doi.org/10.1103/PhysRevB.97.121401} {\bibfield  {journal} {\bibinfo
			{journal} {Physical Review B}\ }\textbf {\bibinfo {volume} {97}},\ \bibinfo
		{pages} {121401(R)} (\bibinfo {year} {2018}{\natexlab{a}})}\BibitemShut
	{NoStop}%
	\bibitem [{\citenamefont {Borgnia}\ \emph {et~al.}(2020)\citenamefont
		{Borgnia}, \citenamefont {Kruchkov},\ and\ \citenamefont
		{Slager}}]{borgniaNonHermitianBoundaryModes2020a}%
	\BibitemOpen
	\bibfield  {author} {\bibinfo {author} {\bibfnamefont {D.~S.}\ \bibnamefont
			{Borgnia}}, \bibinfo {author} {\bibfnamefont {A.~J.}\ \bibnamefont
			{Kruchkov}},\ and\ \bibinfo {author} {\bibfnamefont {R.-J.}\ \bibnamefont
			{Slager}},\ }\bibfield  {title} {\bibinfo {title} {Non-{{Hermitian Boundary
					Modes}} and {{Topology}}},\ }\href
	{https://doi.org/10.1103/PhysRevLett.124.056802} {\bibfield  {journal}
		{\bibinfo  {journal} {Physical Review Letters}\ }\textbf {\bibinfo {volume}
			{124}},\ \bibinfo {pages} {056802} (\bibinfo {year} {2020})}\BibitemShut
	{NoStop}%
	\bibitem [{\citenamefont {Yao}\ and\ \citenamefont
		{Wang}(2018)}]{yaoEdgeStatesTopological2018}%
	\BibitemOpen
	\bibfield  {author} {\bibinfo {author} {\bibfnamefont {S.}~\bibnamefont
			{Yao}}\ and\ \bibinfo {author} {\bibfnamefont {Z.}~\bibnamefont {Wang}},\
	}\bibfield  {title} {\bibinfo {title} {Edge {{States}} and {{Topological
					Invariants}} of {{Non-Hermitian Systems}}},\ }\href
	{https://doi.org/10.1103/PhysRevLett.121.086803} {\bibfield  {journal}
		{\bibinfo  {journal} {Physical Review Letters}\ }\textbf {\bibinfo {volume}
			{121}},\ \bibinfo {pages} {086803} (\bibinfo {year} {2018})}\BibitemShut
	{NoStop}%
	\bibitem [{\citenamefont {Kunst}\ \emph {et~al.}(2018)\citenamefont {Kunst},
		\citenamefont {Edvardsson}, \citenamefont {Budich},\ and\ \citenamefont
		{Bergholtz}}]{kunstBiorthogonalBulkBoundaryCorrespondence2018}%
	\BibitemOpen
	\bibfield  {author} {\bibinfo {author} {\bibfnamefont {F.~K.}\ \bibnamefont
			{Kunst}}, \bibinfo {author} {\bibfnamefont {E.}~\bibnamefont {Edvardsson}},
		\bibinfo {author} {\bibfnamefont {J.~C.}\ \bibnamefont {Budich}},\ and\
		\bibinfo {author} {\bibfnamefont {E.~J.}\ \bibnamefont {Bergholtz}},\
	}\bibfield  {title} {\bibinfo {title} {Biorthogonal {{Bulk-Boundary
					Correspondence}} in {{Non-Hermitian Systems}}},\ }\href
	{https://doi.org/10.1103/PhysRevLett.121.026808} {\bibfield  {journal}
		{\bibinfo  {journal} {Physical Review Letters}\ }\textbf {\bibinfo {volume}
			{121}},\ \bibinfo {pages} {026808} (\bibinfo {year} {2018})}\BibitemShut
	{NoStop}%
	\bibitem [{\citenamefont {Parto}\ \emph {et~al.}(2018)\citenamefont {Parto},
		\citenamefont {Wittek}, \citenamefont {Hodaei}, \citenamefont {Harari},
		\citenamefont {Bandres}, \citenamefont {Ren}, \citenamefont {Rechtsman},
		\citenamefont {Segev}, \citenamefont {Christodoulides},\ and\ \citenamefont
		{Khajavikhan}}]{partoEdgeModeLasing1D2018}%
	\BibitemOpen
	\bibfield  {author} {\bibinfo {author} {\bibfnamefont {M.}~\bibnamefont
			{Parto}}, \bibinfo {author} {\bibfnamefont {S.}~\bibnamefont {Wittek}},
		\bibinfo {author} {\bibfnamefont {H.}~\bibnamefont {Hodaei}}, \bibinfo
		{author} {\bibfnamefont {G.}~\bibnamefont {Harari}}, \bibinfo {author}
		{\bibfnamefont {M.~A.}\ \bibnamefont {Bandres}}, \bibinfo {author}
		{\bibfnamefont {J.}~\bibnamefont {Ren}}, \bibinfo {author} {\bibfnamefont
			{M.~C.}\ \bibnamefont {Rechtsman}}, \bibinfo {author} {\bibfnamefont
			{M.}~\bibnamefont {Segev}}, \bibinfo {author} {\bibfnamefont {D.~N.}\
			\bibnamefont {Christodoulides}},\ and\ \bibinfo {author} {\bibfnamefont
			{M.}~\bibnamefont {Khajavikhan}},\ }\bibfield  {title} {\bibinfo {title}
		{Edge-{{Mode Lasing}} in {{1D Topological Active Arrays}}},\ }\href
	{https://doi.org/10.1103/PhysRevLett.120.113901} {\bibfield  {journal}
		{\bibinfo  {journal} {Physical Review Letters}\ }\textbf {\bibinfo {volume}
			{120}},\ \bibinfo {pages} {113901} (\bibinfo {year} {2018})}\BibitemShut
	{NoStop}%
	\bibitem [{\citenamefont {Bandres}\ \emph {et~al.}(2018)\citenamefont
		{Bandres}, \citenamefont {Wittek}, \citenamefont {Harari}, \citenamefont
		{Parto}, \citenamefont {Ren}, \citenamefont {Segev}, \citenamefont
		{Christodoulides},\ and\ \citenamefont
		{Khajavikhan}}]{bandresTopologicalInsulatorLaser2018}%
	\BibitemOpen
	\bibfield  {author} {\bibinfo {author} {\bibfnamefont {M.~A.}\ \bibnamefont
			{Bandres}}, \bibinfo {author} {\bibfnamefont {S.}~\bibnamefont {Wittek}},
		\bibinfo {author} {\bibfnamefont {G.}~\bibnamefont {Harari}}, \bibinfo
		{author} {\bibfnamefont {M.}~\bibnamefont {Parto}}, \bibinfo {author}
		{\bibfnamefont {J.}~\bibnamefont {Ren}}, \bibinfo {author} {\bibfnamefont
			{M.}~\bibnamefont {Segev}}, \bibinfo {author} {\bibfnamefont {D.~N.}\
			\bibnamefont {Christodoulides}},\ and\ \bibinfo {author} {\bibfnamefont
			{M.}~\bibnamefont {Khajavikhan}},\ }\bibfield  {title} {\bibinfo {title}
		{Topological insulator laser: {{Experiments}}},\ }\bibfield  {journal}
	{\bibinfo  {journal} {Science}\ }\textbf {\bibinfo {volume} {359}},\ \href
	{https://doi.org/10.1126/science.aar4005} {10.1126/science.aar4005} (\bibinfo
	{year} {2018})\BibitemShut {NoStop}%
	\bibitem [{\citenamefont {Harari}\ \emph {et~al.}(2018)\citenamefont {Harari},
		\citenamefont {Bandres}, \citenamefont {Lumer}, \citenamefont {Rechtsman},
		\citenamefont {Chong}, \citenamefont {Khajavikhan}, \citenamefont
		{Christodoulides},\ and\ \citenamefont
		{Segev}}]{harariTopologicalInsulatorLaser2018a}%
	\BibitemOpen
	\bibfield  {author} {\bibinfo {author} {\bibfnamefont {G.}~\bibnamefont
			{Harari}}, \bibinfo {author} {\bibfnamefont {M.~A.}\ \bibnamefont {Bandres}},
		\bibinfo {author} {\bibfnamefont {Y.}~\bibnamefont {Lumer}}, \bibinfo
		{author} {\bibfnamefont {M.~C.}\ \bibnamefont {Rechtsman}}, \bibinfo {author}
		{\bibfnamefont {Y.~D.}\ \bibnamefont {Chong}}, \bibinfo {author}
		{\bibfnamefont {M.}~\bibnamefont {Khajavikhan}}, \bibinfo {author}
		{\bibfnamefont {D.~N.}\ \bibnamefont {Christodoulides}},\ and\ \bibinfo
		{author} {\bibfnamefont {M.}~\bibnamefont {Segev}},\ }\bibfield  {title}
	{\bibinfo {title} {Topological insulator laser: {{Theory}}},\ }\bibfield
	{journal} {\bibinfo  {journal} {Science}\ }\textbf {\bibinfo {volume}
		{359}},\ \href {https://doi.org/10.1126/science.aar4003}
	{10.1126/science.aar4003} (\bibinfo {year} {2018})\BibitemShut {NoStop}%
	\bibitem [{\citenamefont {Albert}\ \emph {et~al.}(2015)\citenamefont {Albert},
		\citenamefont {Glazman},\ and\ \citenamefont
		{Jiang}}]{albertTopologicalPropertiesLinear2015}%
	\BibitemOpen
	\bibfield  {author} {\bibinfo {author} {\bibfnamefont {V.~V.}\ \bibnamefont
			{Albert}}, \bibinfo {author} {\bibfnamefont {L.~I.}\ \bibnamefont
			{Glazman}},\ and\ \bibinfo {author} {\bibfnamefont {L.}~\bibnamefont
			{Jiang}},\ }\bibfield  {title} {\bibinfo {title} {Topological {{Properties}}
			of {{Linear Circuit Lattices}}},\ }\href
	{https://doi.org/10.1103/PhysRevLett.114.173902} {\bibfield  {journal}
		{\bibinfo  {journal} {Physical Review Letters}\ }\textbf {\bibinfo {volume}
			{114}},\ \bibinfo {pages} {173902} (\bibinfo {year} {2015})}\BibitemShut
	{NoStop}%
	\bibitem [{\citenamefont
		{Ezawa}(2019)}]{ezawaNonHermitianHigherorderTopological2019}%
	\BibitemOpen
	\bibfield  {author} {\bibinfo {author} {\bibfnamefont {M.}~\bibnamefont
			{Ezawa}},\ }\bibfield  {title} {\bibinfo {title} {Non-{{Hermitian}}
			higher-order topological states in nonreciprocal and reciprocal systems with
			their electric-circuit realization},\ }\href
	{https://doi.org/10.1103/PhysRevB.99.201411} {\bibfield  {journal} {\bibinfo
			{journal} {Physical Review B}\ }\textbf {\bibinfo {volume} {99}},\ \bibinfo
		{pages} {201411} (\bibinfo {year} {2019})}\BibitemShut {NoStop}%
	\bibitem [{\citenamefont {Liu}\ \emph {et~al.}(2020)\citenamefont {Liu},
		\citenamefont {Ma}, \citenamefont {Yang}, \citenamefont {Zhang},
		\citenamefont {Gao}, \citenamefont {Xiang}, \citenamefont {Cui},\ and\
		\citenamefont {Zhang}}]{liuGainLossInducedTopological2020}%
	\BibitemOpen
	\bibfield  {author} {\bibinfo {author} {\bibfnamefont {S.}~\bibnamefont
			{Liu}}, \bibinfo {author} {\bibfnamefont {S.}~\bibnamefont {Ma}}, \bibinfo
		{author} {\bibfnamefont {C.}~\bibnamefont {Yang}}, \bibinfo {author}
		{\bibfnamefont {L.}~\bibnamefont {Zhang}}, \bibinfo {author} {\bibfnamefont
			{W.}~\bibnamefont {Gao}}, \bibinfo {author} {\bibfnamefont {Y.~J.}\
			\bibnamefont {Xiang}}, \bibinfo {author} {\bibfnamefont {T.~J.}\ \bibnamefont
			{Cui}},\ and\ \bibinfo {author} {\bibfnamefont {S.}~\bibnamefont {Zhang}},\
	}\bibfield  {title} {\bibinfo {title} {Gain- and {{Loss-Induced Topological
					Insulating Phase}} in a {{Non-Hermitian Electrical Circuit}}},\ }\href
	{https://doi.org/10.1103/PhysRevApplied.13.014047} {\bibfield  {journal}
		{\bibinfo  {journal} {Physical Review Applied}\ }\textbf {\bibinfo {volume}
			{13}},\ \bibinfo {pages} {014047} (\bibinfo {year} {2020})}\BibitemShut
	{NoStop}%
	\bibitem [{\citenamefont {Zhao}\ \emph {et~al.}(2018)\citenamefont {Zhao},
		\citenamefont {Miao}, \citenamefont {Teimourpour}, \citenamefont {Malzard},
		\citenamefont {{El-Ganainy}}, \citenamefont {Schomerus},\ and\ \citenamefont
		{Feng}}]{zhaoTopologicalHybridSilicon2018}%
	\BibitemOpen
	\bibfield  {author} {\bibinfo {author} {\bibfnamefont {H.}~\bibnamefont
			{Zhao}}, \bibinfo {author} {\bibfnamefont {P.}~\bibnamefont {Miao}}, \bibinfo
		{author} {\bibfnamefont {M.~H.}\ \bibnamefont {Teimourpour}}, \bibinfo
		{author} {\bibfnamefont {S.}~\bibnamefont {Malzard}}, \bibinfo {author}
		{\bibfnamefont {R.}~\bibnamefont {{El-Ganainy}}}, \bibinfo {author}
		{\bibfnamefont {H.}~\bibnamefont {Schomerus}},\ and\ \bibinfo {author}
		{\bibfnamefont {L.}~\bibnamefont {Feng}},\ }\bibfield  {title} {\bibinfo
		{title} {Topological hybrid silicon microlasers},\ }\href
	{https://doi.org/10.1038/s41467-018-03434-2} {\bibfield  {journal} {\bibinfo
			{journal} {Nature Communications}\ }\textbf {\bibinfo {volume} {9}},\
		\bibinfo {pages} {981} (\bibinfo {year} {2018})}\BibitemShut {NoStop}%
	\bibitem [{\citenamefont {Ozawa}\ \emph {et~al.}(2019)\citenamefont {Ozawa},
		\citenamefont {Price}, \citenamefont {Amo}, \citenamefont {Goldman},
		\citenamefont {Hafezi}, \citenamefont {Lu}, \citenamefont {Rechtsman},
		\citenamefont {Schuster}, \citenamefont {Simon}, \citenamefont {Zilberberg},\
		and\ \citenamefont {Carusotto}}]{ozawaTopologicalPhotonics2019}%
	\BibitemOpen
	\bibfield  {author} {\bibinfo {author} {\bibfnamefont {T.}~\bibnamefont
			{Ozawa}}, \bibinfo {author} {\bibfnamefont {H.~M.}\ \bibnamefont {Price}},
		\bibinfo {author} {\bibfnamefont {A.}~\bibnamefont {Amo}}, \bibinfo {author}
		{\bibfnamefont {N.}~\bibnamefont {Goldman}}, \bibinfo {author} {\bibfnamefont
			{M.}~\bibnamefont {Hafezi}}, \bibinfo {author} {\bibfnamefont
			{L.}~\bibnamefont {Lu}}, \bibinfo {author} {\bibfnamefont {M.~C.}\
			\bibnamefont {Rechtsman}}, \bibinfo {author} {\bibfnamefont {D.}~\bibnamefont
			{Schuster}}, \bibinfo {author} {\bibfnamefont {J.}~\bibnamefont {Simon}},
		\bibinfo {author} {\bibfnamefont {O.}~\bibnamefont {Zilberberg}},\ and\
		\bibinfo {author} {\bibfnamefont {I.}~\bibnamefont {Carusotto}},\ }\bibfield
	{title} {\bibinfo {title} {Topological photonics},\ }\href
	{https://doi.org/10.1103/RevModPhys.91.015006} {\bibfield  {journal}
		{\bibinfo  {journal} {Reviews of Modern Physics}\ }\textbf {\bibinfo {volume}
			{91}},\ \bibinfo {pages} {015006} (\bibinfo {year} {2019})}\BibitemShut
	{NoStop}%
	\bibitem [{\citenamefont {Martinez~Alvarez}\ \emph
		{et~al.}(2018{\natexlab{b}})\citenamefont {Martinez~Alvarez}, \citenamefont
		{Barrios~Vargas}, \citenamefont {Berdakin},\ and\ \citenamefont
		{Foa~Torres}}]{martinezalvarezTopologicalStatesNonHermitian2018}%
	\BibitemOpen
	\bibfield  {author} {\bibinfo {author} {\bibfnamefont {V.~M.}\ \bibnamefont
			{Martinez~Alvarez}}, \bibinfo {author} {\bibfnamefont {J.~E.}\ \bibnamefont
			{Barrios~Vargas}}, \bibinfo {author} {\bibfnamefont {M.}~\bibnamefont
			{Berdakin}},\ and\ \bibinfo {author} {\bibfnamefont {L.~E.~F.}\ \bibnamefont
			{Foa~Torres}},\ }\bibfield  {title} {\bibinfo {title} {Topological states of
			non-{{Hermitian}} systems},\ }\href
	{https://doi.org/10.1140/epjst/e2018-800091-5} {\bibfield  {journal}
		{\bibinfo  {journal} {The European Physical Journal Special Topics}\ }\textbf
		{\bibinfo {volume} {227}},\ \bibinfo {pages} {1295} (\bibinfo {year}
		{2018}{\natexlab{b}})}\BibitemShut {NoStop}%
	\bibitem [{\citenamefont {{\"O}zdemir}\ \emph {et~al.}(2019)\citenamefont
		{{\"O}zdemir}, \citenamefont {Rotter}, \citenamefont {Nori},\ and\
		\citenamefont {Yang}}]{ozdemirParityTimeSymmetry2019}%
	\BibitemOpen
	\bibfield  {author} {\bibinfo {author} {\bibfnamefont {{\c S}.~K.}\
			\bibnamefont {{\"O}zdemir}}, \bibinfo {author} {\bibfnamefont
			{S.}~\bibnamefont {Rotter}}, \bibinfo {author} {\bibfnamefont
			{F.}~\bibnamefont {Nori}},\ and\ \bibinfo {author} {\bibfnamefont
			{L.}~\bibnamefont {Yang}},\ }\bibfield  {title} {\bibinfo {title}
		{Parity\textendash time symmetry and exceptional points in photonics},\
	}\href {https://doi.org/10.1038/s41563-019-0304-9} {\bibfield  {journal}
		{\bibinfo  {journal} {Nature Materials}\ }\textbf {\bibinfo {volume} {18}},\
		\bibinfo {pages} {783} (\bibinfo {year} {2019})},\ \bibinfo {note} {review
		article EP, inclusief simpele hamiltonian}\BibitemShut {NoStop}%
	\bibitem [{\citenamefont {Su}\ \emph {et~al.}(2021)\citenamefont {Su},
		\citenamefont {Estrecho}, \citenamefont {Biega{\'n}ska}, \citenamefont
		{Huang}, \citenamefont {Wurdack}, \citenamefont {Pieczarka}, \citenamefont
		{Truscott}, \citenamefont {Liew}, \citenamefont {Ostrovskaya},\ and\
		\citenamefont {Xiong}}]{suDirectMeasurementNonHermitian2021}%
	\BibitemOpen
	\bibfield  {author} {\bibinfo {author} {\bibfnamefont {R.}~\bibnamefont
			{Su}}, \bibinfo {author} {\bibfnamefont {E.}~\bibnamefont {Estrecho}},
		\bibinfo {author} {\bibfnamefont {D.}~\bibnamefont {Biega{\'n}ska}}, \bibinfo
		{author} {\bibfnamefont {Y.}~\bibnamefont {Huang}}, \bibinfo {author}
		{\bibfnamefont {M.}~\bibnamefont {Wurdack}}, \bibinfo {author} {\bibfnamefont
			{M.}~\bibnamefont {Pieczarka}}, \bibinfo {author} {\bibfnamefont {A.~G.}\
			\bibnamefont {Truscott}}, \bibinfo {author} {\bibfnamefont {T.~C.~H.}\
			\bibnamefont {Liew}}, \bibinfo {author} {\bibfnamefont {E.~A.}\ \bibnamefont
			{Ostrovskaya}},\ and\ \bibinfo {author} {\bibfnamefont {Q.}~\bibnamefont
			{Xiong}},\ }\bibfield  {title} {\bibinfo {title} {Direct measurement of a
			non-{{Hermitian}} topological invariant in a hybrid light-matter system},\
	}\href {https://doi.org/10.1126/sciadv.abj8905} {\bibfield  {journal}
		{\bibinfo  {journal} {Science Advances}\ }\textbf {\bibinfo {volume} {7}},\
		\bibinfo {pages} {eabj8905} (\bibinfo {year} {2021})}\BibitemShut {NoStop}%
	\bibitem [{\citenamefont {Ghatak}\ \emph {et~al.}(2020)\citenamefont {Ghatak},
		\citenamefont {Brandenbourger}, \citenamefont {{\noopsort{wezel}}van Wezel},\
		and\ \citenamefont {Coulais}}]{ghatakObservationNonHermitianTopology2020}%
	\BibitemOpen
	\bibfield  {author} {\bibinfo {author} {\bibfnamefont {A.}~\bibnamefont
			{Ghatak}}, \bibinfo {author} {\bibfnamefont {M.}~\bibnamefont
			{Brandenbourger}}, \bibinfo {author} {\bibfnamefont {J.}~\bibnamefont
			{{\noopsort{wezel}}van Wezel}},\ and\ \bibinfo {author} {\bibfnamefont
			{C.}~\bibnamefont {Coulais}},\ }\bibfield  {title} {\bibinfo {title}
		{Observation of non-{{Hermitian}} topology and its bulk\textendash edge
			correspondence in an active mechanical metamaterial},\ }\href
	{https://doi.org/10.1073/pnas.2010580117} {\bibfield  {journal} {\bibinfo
			{journal} {Proceedings of the National Academy of Sciences}\ }\textbf
		{\bibinfo {volume} {117}},\ \bibinfo {pages} {29561} (\bibinfo {year}
		{2020})}\BibitemShut {NoStop}%
	\bibitem [{\citenamefont {Scheibner}\ \emph {et~al.}(2020)\citenamefont
		{Scheibner}, \citenamefont {Irvine},\ and\ \citenamefont
		{Vitelli}}]{scheibnerNonHermitianBandTopology2020}%
	\BibitemOpen
	\bibfield  {author} {\bibinfo {author} {\bibfnamefont {C.}~\bibnamefont
			{Scheibner}}, \bibinfo {author} {\bibfnamefont {W.~T.~M.}\ \bibnamefont
			{Irvine}},\ and\ \bibinfo {author} {\bibfnamefont {V.}~\bibnamefont
			{Vitelli}},\ }\bibfield  {title} {\bibinfo {title} {Non-{{Hermitian Band
					Topology}} and {{Skin Modes}} in {{Active Elastic Media}}},\ }\href
	{https://doi.org/10.1103/PhysRevLett.125.118001} {\bibfield  {journal}
		{\bibinfo  {journal} {Physical Review Letters}\ }\textbf {\bibinfo {volume}
			{125}},\ \bibinfo {pages} {118001} (\bibinfo {year} {2020})}\BibitemShut
	{NoStop}%
	\bibitem [{\citenamefont {Helbig}\ \emph {et~al.}(2020)\citenamefont {Helbig},
		\citenamefont {Hofmann}, \citenamefont {Imhof}, \citenamefont {Abdelghany},
		\citenamefont {Kiessling}, \citenamefont {Molenkamp}, \citenamefont {Lee},
		\citenamefont {Szameit}, \citenamefont {Greiter},\ and\ \citenamefont
		{Thomale}}]{helbigGeneralizedBulkBoundary2020}%
	\BibitemOpen
	\bibfield  {author} {\bibinfo {author} {\bibfnamefont {T.}~\bibnamefont
			{Helbig}}, \bibinfo {author} {\bibfnamefont {T.}~\bibnamefont {Hofmann}},
		\bibinfo {author} {\bibfnamefont {S.}~\bibnamefont {Imhof}}, \bibinfo
		{author} {\bibfnamefont {M.}~\bibnamefont {Abdelghany}}, \bibinfo {author}
		{\bibfnamefont {T.}~\bibnamefont {Kiessling}}, \bibinfo {author}
		{\bibfnamefont {L.~W.}\ \bibnamefont {Molenkamp}}, \bibinfo {author}
		{\bibfnamefont {C.~H.}\ \bibnamefont {Lee}}, \bibinfo {author} {\bibfnamefont
			{A.}~\bibnamefont {Szameit}}, \bibinfo {author} {\bibfnamefont
			{M.}~\bibnamefont {Greiter}},\ and\ \bibinfo {author} {\bibfnamefont
			{R.}~\bibnamefont {Thomale}},\ }\bibfield  {title} {\bibinfo {title}
		{Generalized bulk\textendash boundary correspondence in non-{{Hermitian}}
			topolectrical circuits},\ }\href {https://doi.org/10.1038/s41567-020-0922-9}
	{\bibfield  {journal} {\bibinfo  {journal} {Nature Physics}\ }\textbf
		{\bibinfo {volume} {16}},\ \bibinfo {pages} {747} (\bibinfo {year}
		{2020})}\BibitemShut {NoStop}%
	\bibitem [{\citenamefont {Lee}\ \emph {et~al.}(2015)\citenamefont {Lee},
		\citenamefont {Kottos},\ and\ \citenamefont
		{Shapiro}}]{leeMacroscopicMagneticStructures2015}%
	\BibitemOpen
	\bibfield  {author} {\bibinfo {author} {\bibfnamefont {J.~M.}\ \bibnamefont
			{Lee}}, \bibinfo {author} {\bibfnamefont {T.}~\bibnamefont {Kottos}},\ and\
		\bibinfo {author} {\bibfnamefont {B.}~\bibnamefont {Shapiro}},\ }\bibfield
	{title} {\bibinfo {title} {Macroscopic magnetic structures with balanced gain
			and loss},\ }\href {https://doi.org/10.1103/PhysRevB.91.094416} {\bibfield
		{journal} {\bibinfo  {journal} {Physical Review B}\ }\textbf {\bibinfo
			{volume} {91}},\ \bibinfo {pages} {094416} (\bibinfo {year}
		{2015})}\BibitemShut {NoStop}%
	\bibitem [{\citenamefont {McClarty}\ and\ \citenamefont
		{Rau}(2019)}]{mcclartyNonHermitianTopologySpontaneous2019}%
	\BibitemOpen
	\bibfield  {author} {\bibinfo {author} {\bibfnamefont {P.~A.}\ \bibnamefont
			{McClarty}}\ and\ \bibinfo {author} {\bibfnamefont {J.~G.}\ \bibnamefont
			{Rau}},\ }\bibfield  {title} {\bibinfo {title} {Non-{{Hermitian}} topology of
			spontaneous magnon decay},\ }\href
	{https://doi.org/10.1103/PhysRevB.100.100405} {\bibfield  {journal} {\bibinfo
			{journal} {Physical Review B}\ }\textbf {\bibinfo {volume} {100}},\ \bibinfo
		{pages} {100405(R)} (\bibinfo {year} {2019})}\BibitemShut {NoStop}%
	\bibitem [{\citenamefont {Deng}\ and\ \citenamefont
		{Flebus}(2021)}]{dengNonHermitianSkinEffect2021}%
	\BibitemOpen
	\bibfield  {author} {\bibinfo {author} {\bibfnamefont {K.}~\bibnamefont
			{Deng}}\ and\ \bibinfo {author} {\bibfnamefont {B.}~\bibnamefont {Flebus}},\
	}\bibfield  {title} {\bibinfo {title} {Non-{{Hermitian}} skin effect in
			magnetic systems},\ }\Eprint {https://arxiv.org/abs/2109.01711}
	{arXiv:2109.01711}  (\bibinfo {year} {2021})\BibitemShut {NoStop}%
	\bibitem [{\citenamefont {Flebus}\ \emph {et~al.}(2020)\citenamefont {Flebus},
		\citenamefont {Duine},\ and\ \citenamefont
		{Hurst}}]{flebusNonHermitianTopologyOnedimensional2020a}%
	\BibitemOpen
	\bibfield  {author} {\bibinfo {author} {\bibfnamefont {B.}~\bibnamefont
			{Flebus}}, \bibinfo {author} {\bibfnamefont {R.~A.}\ \bibnamefont {Duine}},\
		and\ \bibinfo {author} {\bibfnamefont {H.~M.}\ \bibnamefont {Hurst}},\
	}\bibfield  {title} {\bibinfo {title} {Non-{{Hermitian}} topology of
			one-dimensional spin-torque oscillator arrays},\ }\href
	{https://doi.org/10.1103/PhysRevB.102.180408} {\bibfield  {journal} {\bibinfo
			{journal} {Physical Review B}\ }\textbf {\bibinfo {volume} {102}},\ \bibinfo
		{pages} {180408(R)} (\bibinfo {year} {2020})}\BibitemShut {NoStop}%
	\bibitem [{\citenamefont {Slavin}\ and\ \citenamefont
		{Tiberkevich}(2009)}]{slavinNonlinearAutoOscillatorTheory2009}%
	\BibitemOpen
	\bibfield  {author} {\bibinfo {author} {\bibfnamefont {A.}~\bibnamefont
			{Slavin}}\ and\ \bibinfo {author} {\bibfnamefont {V.}~\bibnamefont
			{Tiberkevich}},\ }\bibfield  {title} {\bibinfo {title} {Nonlinear
			{{Auto-Oscillator Theory}} of {{Microwave Generation}} by {{Spin-Polarized
					Current}}},\ }\href {https://doi.org/10.1109/TMAG.2008.2009935} {\bibfield
		{journal} {\bibinfo  {journal} {IEEE Transactions on Magnetics}\ }\textbf
		{\bibinfo {volume} {45}},\ \bibinfo {pages} {1875} (\bibinfo {year}
		{2009})}\BibitemShut {NoStop}%
	\bibitem [{\citenamefont {Kaka}\ \emph {et~al.}(2005)\citenamefont {Kaka},
		\citenamefont {Pufall}, \citenamefont {Rippard}, \citenamefont {Silva},
		\citenamefont {Russek},\ and\ \citenamefont
		{Katine}}]{kakaMutualPhaselockingMicrowave2005}%
	\BibitemOpen
	\bibfield  {author} {\bibinfo {author} {\bibfnamefont {S.}~\bibnamefont
			{Kaka}}, \bibinfo {author} {\bibfnamefont {M.~R.}\ \bibnamefont {Pufall}},
		\bibinfo {author} {\bibfnamefont {W.~H.}\ \bibnamefont {Rippard}}, \bibinfo
		{author} {\bibfnamefont {T.~J.}\ \bibnamefont {Silva}}, \bibinfo {author}
		{\bibfnamefont {S.~E.}\ \bibnamefont {Russek}},\ and\ \bibinfo {author}
		{\bibfnamefont {J.~A.}\ \bibnamefont {Katine}},\ }\bibfield  {title}
	{\bibinfo {title} {Mutual phase-locking of microwave spin torque
			nano-oscillators},\ }\href {https://doi.org/10.1038/nature04035} {\bibfield
		{journal} {\bibinfo  {journal} {Nature}\ }\textbf {\bibinfo {volume} {437}},\
		\bibinfo {pages} {389} (\bibinfo {year} {2005})}\BibitemShut {NoStop}%
	\bibitem [{\citenamefont {Mancoff}\ \emph {et~al.}(2005)\citenamefont
		{Mancoff}, \citenamefont {Rizzo}, \citenamefont {Engel},\ and\ \citenamefont
		{Tehrani}}]{mancoffPhaselockingDoublepointcontactSpintransfer2005}%
	\BibitemOpen
	\bibfield  {author} {\bibinfo {author} {\bibfnamefont {F.~B.}\ \bibnamefont
			{Mancoff}}, \bibinfo {author} {\bibfnamefont {N.~D.}\ \bibnamefont {Rizzo}},
		\bibinfo {author} {\bibfnamefont {B.~N.}\ \bibnamefont {Engel}},\ and\
		\bibinfo {author} {\bibfnamefont {S.}~\bibnamefont {Tehrani}},\ }\bibfield
	{title} {\bibinfo {title} {Phase-locking in double-point-contact
			spin-transfer devices},\ }\href {https://doi.org/10.1038/nature04036}
	{\bibfield  {journal} {\bibinfo  {journal} {Nature}\ }\textbf {\bibinfo
			{volume} {437}},\ \bibinfo {pages} {393} (\bibinfo {year}
		{2005})}\BibitemShut {NoStop}%
	\bibitem [{\citenamefont {Su}\ \emph {et~al.}(1979)\citenamefont {Su},
		\citenamefont {Schrieffer},\ and\ \citenamefont
		{Heeger}}]{suSolitonsPolyacetylene1979}%
	\BibitemOpen
	\bibfield  {author} {\bibinfo {author} {\bibfnamefont {W.~P.}\ \bibnamefont
			{Su}}, \bibinfo {author} {\bibfnamefont {J.~R.}\ \bibnamefont {Schrieffer}},\
		and\ \bibinfo {author} {\bibfnamefont {A.~J.}\ \bibnamefont {Heeger}},\
	}\bibfield  {title} {\bibinfo {title} {Solitons in {{Polyacetylene}}},\
	}\href {https://doi.org/10.1103/PhysRevLett.42.1698} {\bibfield  {journal}
		{\bibinfo  {journal} {Physical Review Letters}\ }\textbf {\bibinfo {volume}
			{42}},\ \bibinfo {pages} {1698} (\bibinfo {year} {1979})},\ \bibinfo {note}
	{sSH chain}\BibitemShut {NoStop}%
	\bibitem [{\citenamefont {Lieu}(2018)}]{lieuTopologicalPhasesNonHermitian2018}%
	\BibitemOpen
	\bibfield  {author} {\bibinfo {author} {\bibfnamefont {S.}~\bibnamefont
			{Lieu}},\ }\bibfield  {title} {\bibinfo {title} {Topological phases in the
			non-{{Hermitian Su-Schrieffer-Heeger}} model},\ }\href
	{https://doi.org/10.1103/PhysRevB.97.045106} {\bibfield  {journal} {\bibinfo
			{journal} {Physical Review B}\ }\textbf {\bibinfo {volume} {97}},\ \bibinfo
		{pages} {045106} (\bibinfo {year} {2018})},\ \bibinfo {note} {topological
		invariants of the non hermitian ssh model}\BibitemShut {NoStop}%
	\bibitem [{\citenamefont {Yokomizo}\ and\ \citenamefont
		{Murakami}(2019)}]{yokomizoNonBlochBandTheory2019}%
	\BibitemOpen
	\bibfield  {author} {\bibinfo {author} {\bibfnamefont {K.}~\bibnamefont
			{Yokomizo}}\ and\ \bibinfo {author} {\bibfnamefont {S.}~\bibnamefont
			{Murakami}},\ }\bibfield  {title} {\bibinfo {title} {Non-{{Bloch Band
					Theory}} of {{Non-Hermitian Systems}}},\ }\href
	{https://doi.org/10.1103/PhysRevLett.123.066404} {\bibfield  {journal}
		{\bibinfo  {journal} {Physical Review Letters}\ }\textbf {\bibinfo {volume}
			{123}},\ \bibinfo {pages} {066404} (\bibinfo {year} {2019})}\BibitemShut
	{NoStop}%
	\bibitem [{\citenamefont
		{Ezawa}(2021)}]{ezawaNonlinearityinducedTransitionNonlinear2021}%
	\BibitemOpen
	\bibfield  {author} {\bibinfo {author} {\bibfnamefont {M.}~\bibnamefont
			{Ezawa}},\ }\bibfield  {title} {\bibinfo {title} {Nonlinearity-induced
			transition in the nonlinear {{Su-Schrieffer-Heeger}} model and a nonlinear
			higher-order topological system},\ }\href
	{https://doi.org/10.1103/PhysRevB.104.235420} {\bibfield  {journal} {\bibinfo
			{journal} {Physical Review B}\ }\textbf {\bibinfo {volume} {104}},\ \bibinfo
		{pages} {235420} (\bibinfo {year} {2021})}\BibitemShut {NoStop}%
	\bibitem [{\citenamefont
		{Ezawa}(2022)}]{ezawaNonlinearNonHermitianHigherorder2022}%
	\BibitemOpen
	\bibfield  {author} {\bibinfo {author} {\bibfnamefont {M.}~\bibnamefont
			{Ezawa}},\ }\bibfield  {title} {\bibinfo {title} {Nonlinear non-{{Hermitian}}
			higher-order topological laser},\ }\href
	{https://doi.org/10.1103/PhysRevResearch.4.013195} {\bibfield  {journal}
		{\bibinfo  {journal} {Physical Review Research}\ }\textbf {\bibinfo {volume}
			{4}},\ \bibinfo {pages} {013195} (\bibinfo {year} {2022})}\BibitemShut
	{NoStop}%
	\bibitem [{\citenamefont {Tiberkevich}\ \emph {et~al.}(2007)\citenamefont
		{Tiberkevich}, \citenamefont {Slavin},\ and\ \citenamefont
		{Kim}}]{tiberkevichMicrowavePowerGenerated2007}%
	\BibitemOpen
	\bibfield  {author} {\bibinfo {author} {\bibfnamefont {V.}~\bibnamefont
			{Tiberkevich}}, \bibinfo {author} {\bibfnamefont {A.}~\bibnamefont
			{Slavin}},\ and\ \bibinfo {author} {\bibfnamefont {J.-V.}\ \bibnamefont
			{Kim}},\ }\bibfield  {title} {\bibinfo {title} {Microwave power generated by
			a spin-torque oscillator in the presence of noise},\ }\href
	{https://doi.org/10.1063/1.2812546} {\bibfield  {journal} {\bibinfo
			{journal} {Applied Physics Letters}\ }\textbf {\bibinfo {volume} {91}},\
		\bibinfo {pages} {192506} (\bibinfo {year} {2007})}\BibitemShut {NoStop}%
	\bibitem [{\citenamefont {Houshang}\ \emph {et~al.}(2016)\citenamefont
		{Houshang}, \citenamefont {Iacocca}, \citenamefont {D{\"u}rrenfeld},
		\citenamefont {Sani}, \citenamefont {{\AA}kerman},\ and\ \citenamefont
		{Dumas}}]{houshangSpinwavebeamDrivenSynchronization2016}%
	\BibitemOpen
	\bibfield  {author} {\bibinfo {author} {\bibfnamefont {A.}~\bibnamefont
			{Houshang}}, \bibinfo {author} {\bibfnamefont {E.}~\bibnamefont {Iacocca}},
		\bibinfo {author} {\bibfnamefont {P.}~\bibnamefont {D{\"u}rrenfeld}},
		\bibinfo {author} {\bibfnamefont {S.~R.}\ \bibnamefont {Sani}}, \bibinfo
		{author} {\bibfnamefont {J.}~\bibnamefont {{\AA}kerman}},\ and\ \bibinfo
		{author} {\bibfnamefont {R.~K.}\ \bibnamefont {Dumas}},\ }\bibfield  {title}
	{\bibinfo {title} {Spin-wave-beam driven synchronization of nanocontact
			spin-torque oscillators},\ }\href {https://doi.org/10.1038/nnano.2015.280}
	{\bibfield  {journal} {\bibinfo  {journal} {Nature Nanotechnology}\ }\textbf
		{\bibinfo {volume} {11}},\ \bibinfo {pages} {280} (\bibinfo {year}
		{2016})}\BibitemShut {NoStop}%
	\bibitem [{\citenamefont {Locatelli}\ \emph {et~al.}(2015)\citenamefont
		{Locatelli}, \citenamefont {Hamadeh}, \citenamefont {Abreu~Araujo},
		\citenamefont {Belanovsky}, \citenamefont {Skirdkov}, \citenamefont {Lebrun},
		\citenamefont {Naletov}, \citenamefont {Zvezdin}, \citenamefont {Mu{\~n}oz},
		\citenamefont {Grollier}, \citenamefont {Klein}, \citenamefont {Cros},\ and\
		\citenamefont {{\noopsort{loubens}}{de
				Loubens}}}]{locatelliEfficientSynchronizationDipolarly2015}%
	\BibitemOpen
	\bibfield  {author} {\bibinfo {author} {\bibfnamefont {N.}~\bibnamefont
			{Locatelli}}, \bibinfo {author} {\bibfnamefont {A.}~\bibnamefont {Hamadeh}},
		\bibinfo {author} {\bibfnamefont {F.}~\bibnamefont {Abreu~Araujo}}, \bibinfo
		{author} {\bibfnamefont {A.~D.}\ \bibnamefont {Belanovsky}}, \bibinfo
		{author} {\bibfnamefont {P.~N.}\ \bibnamefont {Skirdkov}}, \bibinfo {author}
		{\bibfnamefont {R.}~\bibnamefont {Lebrun}}, \bibinfo {author} {\bibfnamefont
			{V.~V.}\ \bibnamefont {Naletov}}, \bibinfo {author} {\bibfnamefont {K.~A.}\
			\bibnamefont {Zvezdin}}, \bibinfo {author} {\bibfnamefont {M.}~\bibnamefont
			{Mu{\~n}oz}}, \bibinfo {author} {\bibfnamefont {J.}~\bibnamefont {Grollier}},
		\bibinfo {author} {\bibfnamefont {O.}~\bibnamefont {Klein}}, \bibinfo
		{author} {\bibfnamefont {V.}~\bibnamefont {Cros}},\ and\ \bibinfo {author}
		{\bibfnamefont {G.}~\bibnamefont {{\noopsort{loubens}}{de Loubens}}},\
	}\bibfield  {title} {\bibinfo {title} {Efficient {{Synchronization}} of
			{{Dipolarly Coupled Vortex-Based Spin Transfer Nano-Oscillators}}},\ }\href
	{https://doi.org/10.1038/srep17039} {\bibfield  {journal} {\bibinfo
			{journal} {Scientific Reports}\ }\textbf {\bibinfo {volume} {5}},\ \bibinfo
		{pages} {17039} (\bibinfo {year} {2015})}\BibitemShut {NoStop}%
	\bibitem [{\citenamefont {Slavin}\ and\ \citenamefont
		{Tiberkevich}(2006)}]{slavinTheoryMutualPhase2006}%
	\BibitemOpen
	\bibfield  {author} {\bibinfo {author} {\bibfnamefont {A.~N.}\ \bibnamefont
			{Slavin}}\ and\ \bibinfo {author} {\bibfnamefont {V.~S.}\ \bibnamefont
			{Tiberkevich}},\ }\bibfield  {title} {\bibinfo {title} {Theory of mutual
			phase locking of spin-torque nanosized oscillators},\ }\href
	{https://doi.org/10.1103/PhysRevB.74.104401} {\bibfield  {journal} {\bibinfo
			{journal} {Physical Review B}\ }\textbf {\bibinfo {volume} {74}},\ \bibinfo
		{pages} {104401} (\bibinfo {year} {2006})}\BibitemShut {NoStop}%
	\bibitem [{\citenamefont {Holstein}\ and\ \citenamefont
		{Primakoff}(1940)}]{holsteinFieldDependenceIntrinsic1940}%
	\BibitemOpen
	\bibfield  {author} {\bibinfo {author} {\bibfnamefont {T.}~\bibnamefont
			{Holstein}}\ and\ \bibinfo {author} {\bibfnamefont {H.}~\bibnamefont
			{Primakoff}},\ }\bibfield  {title} {\bibinfo {title} {Field {{Dependence}} of
			the {{Intrinsic Domain Magnetization}} of a {{Ferromagnet}}},\ }\href
	{https://doi.org/10.1103/PhysRev.58.1098} {\bibfield  {journal} {\bibinfo
			{journal} {Physical Review}\ }\textbf {\bibinfo {volume} {58}},\ \bibinfo
		{pages} {1098} (\bibinfo {year} {1940})}\BibitemShut {NoStop}%
	\bibitem [{\citenamefont {Jin}\ and\ \citenamefont
		{Song}(2019)}]{jinBulkboundaryCorrespondenceNonHermitian2019}%
	\BibitemOpen
	\bibfield  {author} {\bibinfo {author} {\bibfnamefont {L.}~\bibnamefont
			{Jin}}\ and\ \bibinfo {author} {\bibfnamefont {Z.}~\bibnamefont {Song}},\
	}\bibfield  {title} {\bibinfo {title} {Bulk-boundary correspondence in a
			non-{{Hermitian}} system in one dimension with chiral inversion symmetry},\
	}\href {https://doi.org/10.1103/PhysRevB.99.081103} {\bibfield  {journal}
		{\bibinfo  {journal} {Physical Review B}\ }\textbf {\bibinfo {volume} {99}},\
		\bibinfo {pages} {081103(R)} (\bibinfo {year} {2019})}\BibitemShut {NoStop}%
	\bibitem [{\citenamefont {Gong}\ \emph {et~al.}(2016)\citenamefont {Gong},
		\citenamefont {Maghrebi}, \citenamefont {Hu}, \citenamefont {Wall},
		\citenamefont {{Foss-Feig}},\ and\ \citenamefont
		{Gorshkov}}]{gongTopologicalPhasesLongrange2016}%
	\BibitemOpen
	\bibfield  {author} {\bibinfo {author} {\bibfnamefont {Z.-X.}\ \bibnamefont
			{Gong}}, \bibinfo {author} {\bibfnamefont {M.~F.}\ \bibnamefont {Maghrebi}},
		\bibinfo {author} {\bibfnamefont {A.}~\bibnamefont {Hu}}, \bibinfo {author}
		{\bibfnamefont {M.~L.}\ \bibnamefont {Wall}}, \bibinfo {author}
		{\bibfnamefont {M.}~\bibnamefont {{Foss-Feig}}},\ and\ \bibinfo {author}
		{\bibfnamefont {A.~V.}\ \bibnamefont {Gorshkov}},\ }\bibfield  {title}
	{\bibinfo {title} {Topological phases with long-range interactions},\ }\href
	{https://doi.org/10.1103/PhysRevB.93.041102} {\bibfield  {journal} {\bibinfo
			{journal} {Physical Review B}\ }\textbf {\bibinfo {volume} {93}},\ \bibinfo
		{pages} {041102(R)} (\bibinfo {year} {2016})}\BibitemShut {NoStop}%
	\bibitem [{\citenamefont {Kim}\ \emph {et~al.}(2008)\citenamefont {Kim},
		\citenamefont {Mistral}, \citenamefont {Chappert}, \citenamefont
		{Tiberkevich},\ and\ \citenamefont {Slavin}}]{kimLineShapeDistortion2008a}%
	\BibitemOpen
	\bibfield  {author} {\bibinfo {author} {\bibfnamefont {J.-V.}\ \bibnamefont
			{Kim}}, \bibinfo {author} {\bibfnamefont {Q.}~\bibnamefont {Mistral}},
		\bibinfo {author} {\bibfnamefont {C.}~\bibnamefont {Chappert}}, \bibinfo
		{author} {\bibfnamefont {V.~S.}\ \bibnamefont {Tiberkevich}},\ and\ \bibinfo
		{author} {\bibfnamefont {A.~N.}\ \bibnamefont {Slavin}},\ }\bibfield  {title}
	{\bibinfo {title} {Line {{Shape Distortion}} in a {{Nonlinear Auto-Oscillator
					Near Generation Threshold}}: {{Application}} to {{Spin-Torque
					Nano-Oscillators}}},\ }\href {https://doi.org/10.1103/PhysRevLett.100.167201}
	{\bibfield  {journal} {\bibinfo  {journal} {Physical Review Letters}\
		}\textbf {\bibinfo {volume} {100}},\ \bibinfo {pages} {167201} (\bibinfo
		{year} {2008})}\BibitemShut {NoStop}%
	\bibitem [{\citenamefont {Rackauckas}\ and\ \citenamefont
		{Nie}(2017)}]{rackauckasDifferentialEquationsJlPerformant2017}%
	\BibitemOpen
	\bibfield  {author} {\bibinfo {author} {\bibfnamefont {C.}~\bibnamefont
			{Rackauckas}}\ and\ \bibinfo {author} {\bibfnamefont {Q.}~\bibnamefont
			{Nie}},\ }\bibfield  {title} {\bibinfo {title} {{{DifferentialEquations}}.jl
			\textendash{} {{A Performant}} and {{Feature-Rich Ecosystem}} for {{Solving
					Differential Equations}} in {{Julia}}},\ }\href
	{https://doi.org/10.5334/jors.151} {\bibfield  {journal} {\bibinfo  {journal}
			{Journal of Open Research Software}\ }\textbf {\bibinfo {volume} {5}},\
		\bibinfo {pages} {15} (\bibinfo {year} {2017})}\BibitemShut {NoStop}%
	\bibitem [{\citenamefont {Chen}\ \emph {et~al.}(2017)\citenamefont {Chen},
		\citenamefont {Kaya~{\"O}zdemir}, \citenamefont {Zhao}, \citenamefont
		{Wiersig},\ and\ \citenamefont {Yang}}]{chenExceptionalPointsEnhance2017}%
	\BibitemOpen
	\bibfield  {author} {\bibinfo {author} {\bibfnamefont {W.}~\bibnamefont
			{Chen}}, \bibinfo {author} {\bibfnamefont {{\c S}.}~\bibnamefont
			{Kaya~{\"O}zdemir}}, \bibinfo {author} {\bibfnamefont {G.}~\bibnamefont
			{Zhao}}, \bibinfo {author} {\bibfnamefont {J.}~\bibnamefont {Wiersig}},\ and\
		\bibinfo {author} {\bibfnamefont {L.}~\bibnamefont {Yang}},\ }\bibfield
	{title} {\bibinfo {title} {Exceptional points enhance sensing in an optical
			microcavity},\ }\href {https://doi.org/10.1038/nature23281} {\bibfield
		{journal} {\bibinfo  {journal} {Nature}\ }\textbf {\bibinfo {volume} {548}},\
		\bibinfo {pages} {192} (\bibinfo {year} {2017})}\BibitemShut {NoStop}%
	\bibitem [{\citenamefont {Miri}\ and\ \citenamefont
		{Al{\`u}}(2019)}]{miriExceptionalPointsOptics2019}%
	\BibitemOpen
	\bibfield  {author} {\bibinfo {author} {\bibfnamefont {M.-A.}\ \bibnamefont
			{Miri}}\ and\ \bibinfo {author} {\bibfnamefont {A.}~\bibnamefont {Al{\`u}}},\
	}\bibfield  {title} {\bibinfo {title} {Exceptional points in optics and
			photonics},\ }\bibfield  {journal} {\bibinfo  {journal} {Science}\ }\textbf
	{\bibinfo {volume} {363}},\ \href {https://doi.org/10.1126/science.aar7709}
	{10.1126/science.aar7709} (\bibinfo {year} {2019})\BibitemShut {NoStop}%
	\bibitem [{\citenamefont {Dembowski}\ \emph {et~al.}(2004)\citenamefont
		{Dembowski}, \citenamefont {Dietz}, \citenamefont {Gr{\"a}f}, \citenamefont
		{Harney}, \citenamefont {Heine}, \citenamefont {Heiss},\ and\ \citenamefont
		{Richter}}]{dembowskiEncirclingExceptionalPoint2004}%
	\BibitemOpen
	\bibfield  {author} {\bibinfo {author} {\bibfnamefont {C.}~\bibnamefont
			{Dembowski}}, \bibinfo {author} {\bibfnamefont {B.}~\bibnamefont {Dietz}},
		\bibinfo {author} {\bibfnamefont {H.-D.}\ \bibnamefont {Gr{\"a}f}}, \bibinfo
		{author} {\bibfnamefont {H.~L.}\ \bibnamefont {Harney}}, \bibinfo {author}
		{\bibfnamefont {A.}~\bibnamefont {Heine}}, \bibinfo {author} {\bibfnamefont
			{W.~D.}\ \bibnamefont {Heiss}},\ and\ \bibinfo {author} {\bibfnamefont
			{A.}~\bibnamefont {Richter}},\ }\bibfield  {title} {\bibinfo {title}
		{Encircling an exceptional point},\ }\href
	{https://doi.org/10.1103/PhysRevE.69.056216} {\bibfield  {journal} {\bibinfo
			{journal} {Physical Review E}\ }\textbf {\bibinfo {volume} {69}},\ \bibinfo
		{pages} {056216} (\bibinfo {year} {2004})}\BibitemShut {NoStop}%
	\bibitem [{\citenamefont {Vincent}\ \emph {et~al.}(2015)\citenamefont
		{Vincent}, \citenamefont {Larroque}, \citenamefont {Locatelli}, \citenamefont
		{Ben~Romdhane}, \citenamefont {Bichler}, \citenamefont {Gamrat},
		\citenamefont {Zhao}, \citenamefont {Klein}, \citenamefont
		{{Galdin-Retailleau}},\ and\ \citenamefont
		{Querlioz}}]{vincentSpinTransferTorqueMagnetic2015}%
	\BibitemOpen
	\bibfield  {author} {\bibinfo {author} {\bibfnamefont {A.~F.}\ \bibnamefont
			{Vincent}}, \bibinfo {author} {\bibfnamefont {J.}~\bibnamefont {Larroque}},
		\bibinfo {author} {\bibfnamefont {N.}~\bibnamefont {Locatelli}}, \bibinfo
		{author} {\bibfnamefont {N.}~\bibnamefont {Ben~Romdhane}}, \bibinfo {author}
		{\bibfnamefont {O.}~\bibnamefont {Bichler}}, \bibinfo {author} {\bibfnamefont
			{C.}~\bibnamefont {Gamrat}}, \bibinfo {author} {\bibfnamefont {W.~S.}\
			\bibnamefont {Zhao}}, \bibinfo {author} {\bibfnamefont {J.-O.}\ \bibnamefont
			{Klein}}, \bibinfo {author} {\bibfnamefont {S.}~\bibnamefont
			{{Galdin-Retailleau}}},\ and\ \bibinfo {author} {\bibfnamefont
			{D.}~\bibnamefont {Querlioz}},\ }\bibfield  {title} {\bibinfo {title}
		{Spin-{{Transfer Torque Magnetic Memory}} as a {{Stochastic Memristive
					Synapse}} for {{Neuromorphic Systems}}},\ }\href
	{https://doi.org/10.1109/TBCAS.2015.2414423} {\bibfield  {journal} {\bibinfo
			{journal} {IEEE Transactions on Biomedical Circuits and Systems}\ }\textbf
		{\bibinfo {volume} {9}},\ \bibinfo {pages} {166} (\bibinfo {year}
		{2015})}\BibitemShut {NoStop}%
	\bibitem [{\citenamefont {Grollier}\ \emph {et~al.}(2020)\citenamefont
		{Grollier}, \citenamefont {Querlioz}, \citenamefont {Camsari}, \citenamefont
		{{Everschor-Sitte}}, \citenamefont {Fukami},\ and\ \citenamefont
		{Stiles}}]{grollierNeuromorphicSpintronics2020}%
	\BibitemOpen
	\bibfield  {author} {\bibinfo {author} {\bibfnamefont {J.}~\bibnamefont
			{Grollier}}, \bibinfo {author} {\bibfnamefont {D.}~\bibnamefont {Querlioz}},
		\bibinfo {author} {\bibfnamefont {K.~Y.}\ \bibnamefont {Camsari}}, \bibinfo
		{author} {\bibfnamefont {K.}~\bibnamefont {{Everschor-Sitte}}}, \bibinfo
		{author} {\bibfnamefont {S.}~\bibnamefont {Fukami}},\ and\ \bibinfo {author}
		{\bibfnamefont {M.~D.}\ \bibnamefont {Stiles}},\ }\bibfield  {title}
	{\bibinfo {title} {Neuromorphic spintronics},\ }\href
	{https://doi.org/10.1038/s41928-019-0360-9} {\bibfield  {journal} {\bibinfo
			{journal} {Nature Electronics}\ }\textbf {\bibinfo {volume} {3}},\ \bibinfo
		{pages} {360} (\bibinfo {year} {2020})}\BibitemShut {NoStop}%
	\bibitem [{\citenamefont {Hassan}\ \emph {et~al.}(2015)\citenamefont {Hassan},
		\citenamefont {Hodaei}, \citenamefont {Miri}, \citenamefont {Khajavikhan},\
		and\ \citenamefont {Christodoulides}}]{hassanNonlinearReversalMathcalPT2015}%
	\BibitemOpen
	\bibfield  {author} {\bibinfo {author} {\bibfnamefont {A.~U.}\ \bibnamefont
			{Hassan}}, \bibinfo {author} {\bibfnamefont {H.}~\bibnamefont {Hodaei}},
		\bibinfo {author} {\bibfnamefont {M.-A.}\ \bibnamefont {Miri}}, \bibinfo
		{author} {\bibfnamefont {M.}~\bibnamefont {Khajavikhan}},\ and\ \bibinfo
		{author} {\bibfnamefont {D.~N.}\ \bibnamefont {Christodoulides}},\ }\bibfield
	{title} {\bibinfo {title} {Nonlinear reversal of the
			$\mathcal{PT}$-symmetric phase transition in a system of coupled
			semiconductor microring resonators},\ }\href
	{https://doi.org/10.1103/PhysRevA.92.063807} {\bibfield  {journal} {\bibinfo
			{journal} {Physical Review A}\ }\textbf {\bibinfo {volume} {92}},\ \bibinfo
		{pages} {063807} (\bibinfo {year} {2015})}\BibitemShut {NoStop}%
	\bibitem [{\citenamefont {Agrawal}\ and\ \citenamefont
		{Dutta}(1995)}]{agrawalSemiconductorLasers1995}%
	\BibitemOpen
	\bibfield  {author} {\bibinfo {author} {\bibfnamefont {G.~P.}\ \bibnamefont
			{Agrawal}}\ and\ \bibinfo {author} {\bibfnamefont {N.~K.}\ \bibnamefont
			{Dutta}},\ }\href {https://doi.org/10.1007/978-1-4613-0481-4} {\emph
		{\bibinfo {title} {Semiconductor {{Lasers}}}}}\ (\bibinfo  {publisher}
	{{Springer US}},\ \bibinfo {address} {{Boston, MA}},\ \bibinfo {year}
	{1995})\BibitemShut {NoStop}%
	\bibitem [{\citenamefont {Ding}\ and\ \citenamefont
		{Miri}(2019)}]{dingModeDiscriminationDissipatively2019}%
	\BibitemOpen
	\bibfield  {author} {\bibinfo {author} {\bibfnamefont {J.}~\bibnamefont
			{Ding}}\ and\ \bibinfo {author} {\bibfnamefont {M.-A.}\ \bibnamefont
			{Miri}},\ }\bibfield  {title} {\bibinfo {title} {Mode discrimination in
			dissipatively coupled laser arrays},\ }\href
	{https://doi.org/10.1364/OL.44.005021} {\bibfield  {journal} {\bibinfo
			{journal} {Optics Letters}\ }\textbf {\bibinfo {volume} {44}},\ \bibinfo
		{pages} {5021} (\bibinfo {year} {2019})}\BibitemShut {NoStop}%
	\bibitem [{\citenamefont {Slavin}(2009)}]{slavinSpintorqueOscillatorsGet2009}%
	\BibitemOpen
	\bibfield  {author} {\bibinfo {author} {\bibfnamefont {A.}~\bibnamefont
			{Slavin}},\ }\bibfield  {title} {\bibinfo {title} {Spin-torque oscillators
			get in phase},\ }\href {https://doi.org/10.1038/nnano.2009.213} {\bibfield
		{journal} {\bibinfo  {journal} {Nature Nanotechnology}\ }\textbf {\bibinfo
			{volume} {4}},\ \bibinfo {pages} {479} (\bibinfo {year} {2009})}\BibitemShut
	{NoStop}%
	\bibitem [{\citenamefont {Wittrock}\ \emph {et~al.}(2021)\citenamefont
		{Wittrock}, \citenamefont {Perna}, \citenamefont {Lebrun}, \citenamefont
		{Dutra}, \citenamefont {Ferreira}, \citenamefont {Bortolotti}, \citenamefont
		{Serpico},\ and\ \citenamefont
		{Cros}}]{wittrockExceptionalPointsControlling2021}%
	\BibitemOpen
	\bibfield  {author} {\bibinfo {author} {\bibfnamefont {S.}~\bibnamefont
			{Wittrock}}, \bibinfo {author} {\bibfnamefont {S.}~\bibnamefont {Perna}},
		\bibinfo {author} {\bibfnamefont {R.}~\bibnamefont {Lebrun}}, \bibinfo
		{author} {\bibfnamefont {R.}~\bibnamefont {Dutra}}, \bibinfo {author}
		{\bibfnamefont {R.}~\bibnamefont {Ferreira}}, \bibinfo {author}
		{\bibfnamefont {P.}~\bibnamefont {Bortolotti}}, \bibinfo {author}
		{\bibfnamefont {C.}~\bibnamefont {Serpico}},\ and\ \bibinfo {author}
		{\bibfnamefont {V.}~\bibnamefont {Cros}},\ }\bibfield  {title} {\bibinfo
		{title} {Exceptional points controlling oscillation death in coupled
			spintronic nano-oscillators},\ }\Eprint {https://arxiv.org/abs/2108.04804}
	{arXiv:2108.04804}  (\bibinfo {year} {2021})\BibitemShut {NoStop}%
\end{thebibliography}

\newcommand{\noopsort}[1]{}

\end{document}